\documentclass[final,3p,times,sort&compress,twocolumn]{elsarticle}

\usepackage{epsfig}
\usepackage{amsthm}
\usepackage{amsmath,amssymb}
\usepackage{color}
\usepackage{graphicx}
\usepackage[table]{xcolor}
\usepackage{lscape}
\usepackage{caption}
\usepackage{longtable}
\usepackage{soul}
\journal{Physica E}

\begin{document}

\begin{frontmatter}



\title{Influence of applied electric and magnetic fields on a thermally-induced reentrance of a coupled spin-electron model on a decorated square lattice}


%%%\author{}
\author{Hana \v Cen\v carikov\'a\corref{cor1}\fnref{label1}}
\cortext[cor1]{Corresponding author:}
\ead{hcencar@saske.sk}
\author{Jozef Stre\v{c}ka\fnref{label2}}
\author{Andrej Gendiar\fnref{label3}}

\address[label1]{Institute of Experimental Physics, Slovak Academy of Sciences, Watsonova 47, 040 01 Ko\v {s}ice, Slovakia}
\address[label2]{Department of Theoretical Physics and Astrophysics, Faculty of Science, P.~J. \v{S}af\' arik University, Park Angelinum 9, 040 01 Ko\v{s}ice, Slovakia}
\address[label3]{Institute of Physics, Slovak Academy of Sciences, D\'{u}bravsk\'{a} cesta 9, SK-845 11, Bratislava, Slovakia}
\begin{abstract}
 The combination of an exact and Corner Transfer Matrix Renormalization Group (CTMRG) methods is used to study an influence of external electric and magnetic fields on existence of intriguing reentrant magnetic transitions in a coupled spin-electron model on a decorated square lattice. The two-dimensional (2D) decorated square lattice with localized nodal spins and delocalized electrons is taken into account. It was found that the competition among all involved interactions (the electron  hopping, spin-spin and spin-electron interaction, external electric and magnetic fields) in combination with thermal fluctuations can produce new type of reentrant magnetic transitions. Depending on the model parameters the non-zero fields can stabilize or destabilize magnetic reentrance. In addition, an alternative and more effective way, for modulating the magnetic reentrance is found. An origin of intriguing low-temperature round maximum in the specific heat was explained as a consequence of rapid changes in the sublattice magnetizations, which is induced through a competition of all presented interactions.  
\end{abstract}


\end{frontmatter}


\section{Introduction}
\label{s1}

The research in the condensed matter physics, material sciences and engineering is characterized  during the last decades by the extensive studies with a goal to synthesize and to apprehend the novel materials, which could be implemented in the technology process with less energy demands. In addition, it is expected that such materials  should be environmentally  friendly and useful for constructing of new devices for our daily life. Indispensable criterion for such novel materials is being their proportions (dimensions), when taking into account the tendency to design facilities smaller in each direction. 
The possible candidates, which satisfy aforementioned requirements,  can be found in a wide group of materials known as correlated spin-electron systems, many of which is intrinsically low-dimensional. The correlated spin-electron systems are, in principle, the layered or quasi-low-dimensional systems with a variety of unconventional structural, magnetic, electronic and transport properties~\cite{Kanamori,Takada,Honecker,Kikuchi,Kamihara1,Kamihara2,Li,Koster}.
An exhaustive study in this research field disclosed that  presence of  exotic cooperative phenomena has its origin dominantly in a competition between magnetic and electronic subsystems~\cite{Velu,Baibich}. However, there exist numerous studies, e.g.~\cite{Motrunich,Cenci3a,Liu}, which  show on the importance of further interactions in such defined systems and the correct microscopic model as well as microscopic mechanisms are still highly debate problems. 
The one possible reason why this problem still resists  our understanding is connected to a huge system complexity, where the widely cooperative nature and many active degrees of freedom make such materials almost intractable for the most analytical theories. There arise alternative models and methods~\cite{Falicov,Anderson,Pereira1, Pereira2,Lisnii1,Lisnii2,Lanczos, Dagotto,Malvezzi,White1,White2,Newman,MC2,Engel}, which allow to transform the complex problem to  simpler counterpart. Among them one can find a very appealing approach based on the  Fisher's mapping idea~\cite{Fisher}, where an arbitrary statistical-mechanical system  interacting purely with the Ising spins may be replaced by new effective 
interactions between the Ising spins using the generalized mapping transformations~\cite{Fisher,Syozi}. Despite of the method simplicity, the application of this method allowed to get many interesting results for various correlated spin-electron systems in one~\cite{Pereira1, Pereira2, Cisarova,Cisarova2,Cisarova3,Galisova3,Torrico,Galisova2017,Carvalho,Galisova2018,Sousa} or two dimensions~\cite{Galisova,Galisova2,Doria,Cenci1,Strecka2009} with a good correspondence between theory and experiments.

In this paper we will concentrate our attention to a comprehensive analysis of a coupled spin-electron model on a decorated square lattice under the simultaneous influence of external magnetic and electric fields, with the main emphasis laid on a stability of reentrant magnetic transitions. The reentrant magnetic transition is a fascinating  phenomenon detected in various complex systems, such as superconductors~\cite{Fisher}, spin glasses~\cite{Binder,Maletta,Fisher2} or intermetallic compounds~\cite{Kolmakova,Venturini,Welter} and it denotes re-appearance of one and the same magnetic phase with  identical or very  similar properties  mediated by the phase transition through the another magnetic phase. The full  understanding of the reentrant phenomenon is highly desirable, since such behavior is connected to a huge application potential in sensors or detectors, when small changes of  temperature or external fields can drive the magnetic state.    
At present it was found that the type of magnetic reentrance can be modified through  the chemical composition~\cite{Venturini,Welter}, which alters various types and/or sizes of underlying interactions~\cite{Cenci2}. To generate  the  reentrance with required properties, e.g. with a manifold of consecutive critical points, is however very complicated and demanding process. We suppose that the magnetoelectric character of a coupled spin-electron system provides  an alternative way for a control of magnetic reentrance through the application of external magnetic and/or electric field, which allow us to change magnetic properties as well as the type of reentrance in a rather straightforward way. 

The outline of this papers is as follows. In Sec.~2 we will briefly describe the investigated model and derive all  physical quantities necessary for our next studies. The most interesting  results will be presented  in Sec.~3, and finally, some concluding remarks will be draw in Sec.~4.

\section{Model and Method}
\label{s2}
\begin{figure}[h!]
\begin{center}
{\includegraphics[scale=0.65,trim=0cm 0cm 0cm 0cm, clip]{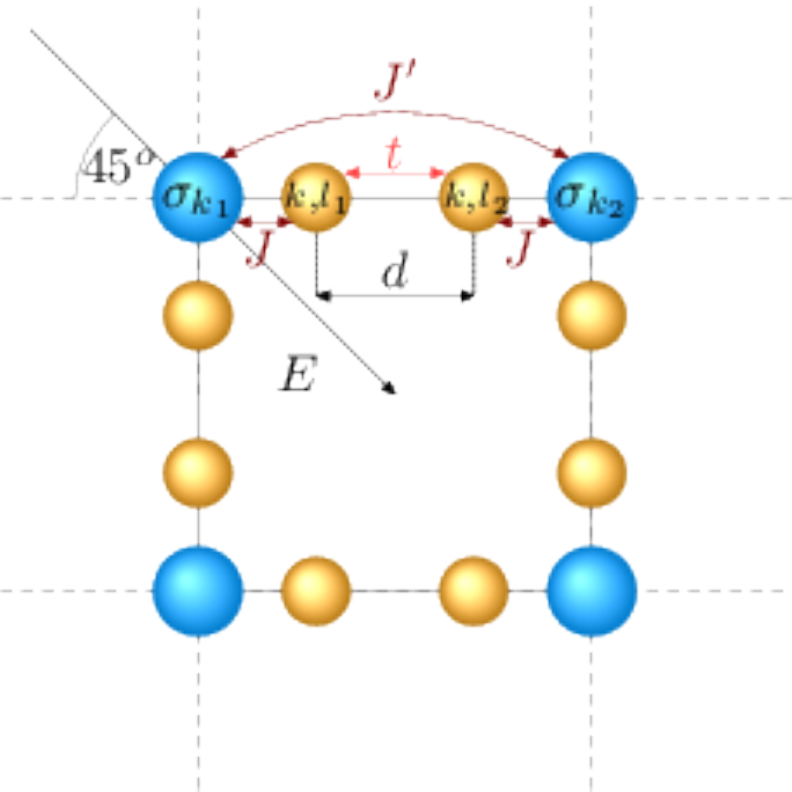}}
\caption{\small  A schematic representation of a small fragment of the studied spin-electron model (\ref{eq1}) on a doubly decorated square lattice. Bigger balls correspond to nodal lattice sites occupied by the localized Ising spins, while smaller balls to decorating sites occupied by at most four mobile electrons per dimer. The interactions assumed within the $k$-th bond are visualized.}
\label{fig1}
\end{center}
\end{figure}

Our analysis is restricted to the interacting spin-electron model on a doubly decorated  square lattice, where the localized Ising spins are placed at  nodal points of a square lattice and the mobile electrons are delocalized over the pairs of decorating sites (dimers) placed at each bond between two nearest-neighbour spins (Fig.~\ref{fig1}). 
The Hamiltonian $\hat{\cal{H}}\!=\!\sum_{k=1}^{2N}\hat{\cal{H}}_k$ of an investigated spin-electron system can be introduced as a sum of the bond Hamiltonians  $\hat{\cal{H}}_k$, given by 
\allowdisplaybreaks
\begin{eqnarray}
\hspace*{-3cm}
\hat{\cal H}_k\!\!\!&=&\!\!\!-t(\hat{c}^\dagger_{k,l_1,\uparrow}\hat{c}_{k,l_2,\uparrow}\!+\!\hat{c}^\dagger_{k,l_1,\downarrow}\hat{c}_{k,l_2,\downarrow}\!+\!
h.c.)\!-\!J\hat\sigma^z_{k_1}(\hat{n}_{k,l_1,\uparrow}
\nonumber
\\
\!\!\!&-&\!\!\!\hat{n}_{k,l_1,\downarrow})-
J\hat\sigma^z_{k_2}(\hat{n}_{k,l_2,\uparrow}\!-\!\hat{n}_{k,l_2,\downarrow})-J'\hat\sigma^z_{k_1}\hat\sigma^z_{k_2}
\nonumber\\
\!\!\!&-&\!\!\!h(\hat{n}_{k,l_1,\uparrow}\!-\!\hat{n}_{k,l_1,\downarrow})\!-\!
h(\hat{n}_{k,l_2,\uparrow}\!-\!\hat{n}_{k,l_2,\downarrow})
\!-\!\frac{h}{4}(\hat\sigma^z_{k_1}\!+\!\hat\sigma^z_{k_2})
\nonumber\\
\!\!\!&-&\!\!\!\frac{V}{\sqrt{2}}(\hat{n}_{k,l_1}\!-\!\hat{n}_{k,l_2})\!-\!\mu \hat{n}_{k}.
\label{eq1}
\end{eqnarray}
Here, the first term  corresponds to the kinetic energy of mobile electrons fluctuating between two (decorating) lattice sites $l_1$ and $l_2$ at each $k$-th bond, the symbols  $\hat{c}^\dagger_{k,l_{\alpha},\gamma}$ ($\hat{c}_{k,l_{\alpha},\gamma}$), ($\alpha\!=\!1$,2; $\gamma\!=\!\uparrow,\downarrow$) are used to denote the creation (annihilation) fermionic operators with their respective number operator $\hat{n}_{k,l_{\alpha},\gamma}\!=\!\hat{c}^\dagger_{k,l_{\alpha},\gamma}\hat{c}_{k,l_{\alpha},\gamma}$. The spin-electron interaction between the adjacent Ising spins $\sigma=\pm 1$ and the mobile electrons is modulated through the next two terms with the coupling constant $J$, while the further-neighbour interaction between localized spins from the same bond is described through the coupling constant $J'$. The further-neighbour interaction can be of either  ferromagnetic ($J'\!>\!0$) or antiferromagnetic ($J'\!<\!0$) one. The application of an external magnetic field $h$ is implemented through the terms in the third line of Eq.~(\ref{eq1}) and acts equally on the localized Ising spins as well as mobile electrons. The external electric field $E$ acts via the electrostatic potential $V\!=\!E|e|d/2$ ($e$ is an electron charge and $d$ is a distance between decorating atoms) on the mobile electrons only. In the first approximation we concentrate only on the simplest case, where the angle between the electric field ${E}$ and the electron-pair distance ${d}$ has a fixed value of $\pi/4$. In such special case, the energy contributions originating from the horizontal and vertical bonds are identical, i.e. $V/\sqrt{2}$. Finally, the last term in Eq.~(\ref{eq1}) allows to control the number of mobile electrons  by their chemical potential $\mu$.
Due to the local character of all assumed interactions, the generalized decoration-iteration transformation~\cite{Fisher,Syozi,Rojas} can be used to obtain an exact expression for the grand-canonical partition function $\Xi$ factorized into a product of the partial bond partition functions $\Xi_k$
\begin{eqnarray}
\Xi\!\!\!&=&\!\!\!\sum_{\{\sigma\}}\prod_{k=1}^{2N}\mbox{Tr}_k\exp(-\beta\hat{\cal H}_k)=\sum_{\{\sigma\}}\prod_{k=1}^{2N}\Xi_k.
\label{eq2}
\end{eqnarray}
Here, $\beta\!=\!1/k_BT$, $k_B$ is  Boltzmann's constant, $T$ is the absolute temperature, the summation $\sum_{\{\sigma\}}$ runs over all configurations of Ising spins $\{\sigma\}$ and the trace ${\rm Tr}_k$ is performed over degrees of the freedom of mobile electrons from the $k$-th bond only. For this reason, the bond partition function $\Xi_k$ can be exclusively converted to the function solely depending on the states of the Ising spins.  A direct consequence of this mathematical transformation is the substitution of a more complex problem to its simpler counterpart. In this case, the generalized decoration-iteration transformation enables to solve instead of the complicated spin-electron system the novel purely localized  Ising spin model on an undecorated (square) lattice given by new effective parameters  $A$, $R$ and $h_{\rm eff}$ 
\begin{eqnarray}
\Xi_k=A\exp(\beta R\sigma_{k_1}\sigma_{k_2})\exp\left(\frac{\beta h_{\rm eff}(\sigma_{k_1}\!+\!\sigma_{k_2})}{4}\right).
\label{eq3} 
\end{eqnarray}
Substituting the transformation relation~(\ref{eq3}) into the expression~(\ref{eq2}), one can obtain the simple mapping relation between the grand-canonical partition function $\Xi$ of a decorated spin-electron model under the investigation and the canonical partition function ${\cal Z}_{IM}$ of an undecorated effective Ising model 
\begin{eqnarray}
\Xi(\beta,J,t,V,h)=A^{2N}{\cal Z}_{IM}(\beta,R,h_{eff}).
\label{eq6a}
\end{eqnarray}
The expressions for the mapping parameters  ($A$, $R$ and $h_{\rm eff}$) are given self-consistently by Eq.~(\ref{eq3}), which must be hold  for all configurations of Ising spins $\sigma_{k_1}$ and $\sigma_{k_2}$. Then, one gets
\begin{eqnarray}
A\!\!\!&=&\!\!\!(V_1V_2V_3V_4)^{1/4},\hspace*{0.5cm} \beta R=\frac{1}{4}\ln\left(\frac{V_1V_2}{V_3V_4}\right),\nonumber\\
\beta h_{\rm eff}\!\!\!&=&\!\!\!\ln\left(\frac{V_1}{V_2}\right),
\label{eq4}
\end{eqnarray}
where 
\allowdisplaybreaks
\begin{eqnarray}
V_1\!\!\!&=&\!\!\!e^{\beta\left(J'+h/2\right)}\left\{1\!+\!4(z+z^3)\cosh \beta\left(J\!+\!h\right)\cosh\beta t^*
\right.
\nonumber\\\!\!\!&+&\!\!\!\left.
2z^2\left[ 1\!+\!\cosh2\beta\left(J\!+\!h\right)\!+\!
\cosh2\beta t^*\right]\!+\!z^4\right\},
\nonumber\\
V_2\!\!\!&=&\!\!\!e^{\beta\left(J'-h/2\right)}\left\{1\!+\!4(z\!+\!z^3)\cosh \beta\left(J\!-\!h\right)\cosh\beta t^*\right.
\nonumber\\\!\!\!&+&\!\!\!\left.
2z^2\left[ 1\!+\!\cosh2\beta\left(J\!-\!h\right)\!+\!
\cosh2\beta t^*\right]\!+\!z^4\right\},
\nonumber\\
V_3\!\!\!&=&\!\!\!e^{-J'\beta}\left\{1\!+\!2z\left[\frac{\cosh\beta {B_+}}{e^{-\beta h}}\!+\!\frac{\cosh\beta {B_{-}}}{e^{\beta h}} \right]\right.
\nonumber\\\!\!\!&+&\!\!\!\left.
2z^3\left[\frac{\cosh\beta {B_+}}{e^{\beta h}}\!+\!\frac{\cosh\beta {B_{-}}}{e^{-\beta h}}\right]\!+\!z^4\right.
\nonumber\\
\!\!\!&+&\!\!\!\left. 2z^2\left[\cosh 2\beta h\!+\!\cosh\beta {F_+}\!+\!\cosh\beta {F_-}\right]\right\},\nonumber\\
V_4\!\!\!&=&\!\!\!e^{-J'\beta}\left\{1+2z\left[\frac{\cosh\beta {B_+}}{e^{\beta h}}\!+\!\frac{\cosh\beta {B_-}}{e^{-\beta h}}\right]\right.
\nonumber\\\!\!\!&+&\!\!\!\left.
2z^3\left[\frac{\cosh\beta {B_+}}{e^{-\beta h}}\!+\!\frac{\cosh\beta {B_-}}{e^{\beta h}}\right]\!+\!z^4\right.
\nonumber\\
\!\!\!&+&\!\!\!\left. 2z^2\left[\cosh 2\beta h\!+\!\cosh\beta {F_+}\!+\!\cosh\beta {F_-}\right]\right\}.
\label{eq5}
\end{eqnarray}
Here, we have defined following expressions 
\begin{eqnarray}
t^*\!\!\!&=&\!\!\!\sqrt{{V^2}/2+t^2},\;\;\; {B}_{\pm}=\sqrt{J[J\pm\sqrt{2}V]+{t^*}^2},\;\;\nonumber\\ 
{F}_{\pm}\!\!\!&=&\!\!\!\sqrt{2\left[J^2+{t^*}^2\right]\pm2
{B}_+{B}_-},\;\;\;z=e^{\beta\mu}. 
\label{eq6}
\end{eqnarray}

A crucial quantity for the following physical analyses is the average number of mobile electrons $\langle \hat{n}_{k}\rangle\equiv\rho$ per bond, which represents an equation of state for a coupled spin-electron model on a decorated square lattice (\ref{eq1}).
\allowdisplaybreaks
\begin{eqnarray}
\rho\!\!\!&=&\!\!\!\frac{z}{2N}\frac{\partial \ln \Xi}{\partial z}\!=\!
z\frac{\partial \ln A}{\partial z}+z\varepsilon\frac{\partial \beta R}{\partial z}+\frac{z}{2}\langle \sigma_i\rangle\frac{\partial \beta h_{eff}}{\partial z}
\nonumber\\
\!\!\!&=&\!\!\!\frac{z}{4}\left\{\frac{V'_{1}}{V_1}\left(1+\varepsilon+\frac{\langle \sigma_i\rangle}{2}\right)+\frac{V'_{2}}{V_2}\left(1+\varepsilon-\frac{\langle \sigma_i\rangle}{2}\right)\right.
\nonumber\\
\!\!\!&+&\!\!\!\left.\left(1-\varepsilon\right)\left(\frac{V'_{3}}{V_3}+\frac{V'_{4}}{V_4}\right)\right\}\!.
\label{eq7}
\end{eqnarray}
Here, the symbol $\varepsilon=\langle\sigma_{k_1}\sigma_{k_2}\rangle$ denotes the pair-correlation function between the localized Ising spins and $V_1'\!=\!\frac{\partial V_1}{\partial z}$, $V_2'\!=\!\frac{\partial V_2}{\partial z}$.
For our analyses focusing onto the study of magnetic reentrant transitions is necessary to define the uniform magnetization of the localized Ising spins $m_i$ and mobile electrons $m_e$ as possible order parameters describing the ferromagnetic (F) spontaneous long-range order at each subsystems, 
\begin{eqnarray}
\hspace*{-0.5cm}
m_{i}\!\!\!\!\!&=&\!\!\!\!\!\frac{1}{2}\langle \langle{\sigma}_{k_1}\rangle\!+\!\langle{\sigma}_{k_2}\rangle\rangle=m_{IM},
\label{eq8}\\
m_{e}\!\!\!\!\!&=&\!\!\!\!\!\Bigg\langle\! \frac{1}{\Xi_k}\left[\frac{\partial \Xi_k}{\partial \beta J\sigma_{k_1}}\!+\!\frac{\partial \Xi_k}{\partial \beta J\sigma_{k_2}}\right]\!\Bigg\rangle
\!=\!\frac{m_i}{2}\left(\frac{W_1}{V_1}\!-\!\frac{W_2}{V_2}\right)\nonumber\\
\!\!\!\!\!&+&\!\!\!
\!\!\frac{1\!+\!\varepsilon}{4}\left(\frac{W_1}{V_1}\!+\!\frac{W_2}{V_2}\!+\!\frac{W_3}{V_3}\!+\!\frac{W_4}{V_4}\right)\!-\!\frac{\varepsilon}{2}\left(\frac{W_3}{V_3}\!+\frac{W_4}{V_4}\right).
\nonumber
\end{eqnarray}
For brevity, we have introduced here new functions
\begin{eqnarray}
{W_1}\!\!\!&=&\!\!\!4e^{\beta(J'+h/2)}\left\{(z\!+\!z^3)\sinh\beta\left(J\!+\!h\right)\cosh\beta t^*
\right.\nonumber\\\!\!\!&+&\!\!\!\left.
z^2\sinh 2\beta (J\!+\!h)\right\},\nonumber\\
{W_2}\!\!\!&=&\!\!\!-4e^{\beta(J'-h/2)}\left\{(z\!+\!z^3)\sinh\beta\left(J\!-\!h\right)\cosh\beta t^*
\right.\nonumber\\\!\!\!&+&\!\!\!\left.
z^2\sinh 2\beta (J\!-\!h)\right\},\nonumber\\
{W_3}\!\!\!&=&\!\!\!2e^{-J'\beta}\left\{z\left[ \frac{\cosh\beta {B_+}}{e^{-\beta h}}\!-\!\frac{\cosh\beta {B_-}}{e^{\beta h}}\right]
\right.\nonumber\\\!\!\!&+&\!\!\!\left.
\!\!\!z^3\left[\frac{\cosh\beta {B_-}}{e^{-\beta h}}\!-\!\frac{\cosh\beta {B_+}}{e^{\beta h}}\right]\!+\!2z^2\sinh2\beta h\right\},\nonumber\\
{W_4}\!\!\!&=&\!\!\!2e^{-J'\beta}\left\{z\left[\frac{\cosh\beta {B_-}}{e^{-\beta h}}\!-\!\frac{\cosh\beta {B_+}}{-e^{\beta h}}\right]
\right.\label{eq10}\\\!\!\!&+&\!\!\!\left.
\!\!\!z^3\left[\frac{\cosh\beta {B_+}}{e^{-\beta h}}\!-\!\frac{\cosh\beta {B_-}}{e^{\beta h}}\right]\!+\!2z^2\sinh2\beta h\right\}.
\nonumber
\end{eqnarray}
For the characterization of the antiferromagnetic (AF) spontaneous long-range order at each subsystems we define another order parameters, namely, the staggered magnetization of the localized Ising spins $m_i^s$ and mobile electrons $m_e^s$ as a difference  of two nearest-neighbour magnetic moments at the selected bond
\begin{eqnarray}
m_{i}^s\!\!\!&=&\!\!\!\frac{1}{2}\langle \langle{\sigma}_{k_1}\rangle\!-\!\langle{\sigma}_{k_2}\rangle\rangle=m_{IM}^s,
\label{eq11}\\
m_{e}^s\!\!\!&=&\!\!\!\Bigg\langle \frac{1}{\Xi_k}\left[\frac{\partial \Xi_k}{\partial \beta J\sigma_{k_1}}\!-\!\frac{\partial \Xi_k}{\partial \beta J\sigma_{k_2}}\right]\Bigg\rangle
\nonumber\\
\!\!\!&=&\!\!\!\frac{(1-\varepsilon)}{4}\left(\frac{T_1}{V_1}\!+\!\frac{T_2}{V_2}\!+\!\frac{T_3}{V_3}\!+\!\frac{T_4}{V_4}\right)\!+\!\frac{\varepsilon}{2}\left(\frac{T_1}{V_1}\!+\!\frac{T_2}{V_2}\right)
\nonumber\\
\!\!\!&+&\!\!\!
\frac{\left(m_i\!+\!m_i^s\right)}{2}\left(\frac{T_1}{V_1}\!-\!\frac{T_2}{V_2}\right)
\!+\!\frac{m_i^s}{2}\left(\frac{T_3}{V_3}\!-\!\frac{T_4}{V_4}\right).
\label{eq12}
\end{eqnarray}
In this case, the staggered magnetization~(\ref{eq12}) is expressed in terms of the following functions 
\allowdisplaybreaks
\begin{eqnarray}
{T_1}\!\!\!&=&\!\!\!\frac{4V}{\sqrt{2}t^*}(z\!-\!z^3)e^{\beta(J'+2h/q)}\sinh\beta (J\!+\!h)\sinh\beta t^*,
\nonumber\\
 {T_2}\!\!\!&=&\!\!\!\frac{-4V}{\sqrt{2}t^*}(z\!-\!z^3)e^{\beta(J'-2h/q)}\sinh\beta (J\!-\!h)\sinh\beta t^*,
\nonumber\\
{T_3}\!\!\!&=&\!\!\! 2e^{-J'\beta}\left\{ z\left[ \frac{\Omega_{+}}{e^{-\beta h}}\frac{\sinh\beta B_+}{B_+}+ \frac{\Omega_{-}}{e^{\beta h}}\frac{\sinh\beta B_-}{B_-}\right]
\right.
\nonumber\\
\!\!\!&+&\!\!\! z^3\left[ \frac{\Omega_{+}}{e^{\beta h}}\frac{\sinh\beta B_+}{B_+}+ \frac{\Omega_{-}}{e^{-\beta h}}\frac{\sinh\beta B_-}{B_-}\right]
%\right.
\nonumber\\
\!\!\!&+&\!\!\!z^2\left[\frac{\sinh \beta F_+}{F_+}\left(2J+ \frac{\Omega_{+}B_-}{B_+}+ \frac{\Omega_{-}B_+}{B_-}  \right) \right.
\nonumber\\
\!\!\!&+&\!\!\! \left.\left.
\frac{\sinh \beta F_-}{F_-}\left(2J+ \frac{\Omega_{-}B_-}{B_+}+ \frac{\Omega_{+}B_+}{B_-}  \right)\right]\right\},\nonumber\\
{T_4}\!\!\!&=&\!\!\! -2e^{-J'\beta}\left\{ z\left[\frac{\Omega_{+}}{e^{\beta h}}\frac{\sinh\beta B_+}{B_+}+ \frac{\Omega_{-}}{e^{-\beta h}}\frac{\sinh\beta B_-}{B_-}\right]
\right.
\nonumber\\
\!\!\!&+&\!\!\! z^3\left[  \frac{\Omega_{+}}{e^{-\beta h}}\frac{\sinh\beta B_+}{B_+}+\frac{\Omega_{-}}{e^{\beta h}}\frac{\sinh\beta B_-}{B_-}\right]
%\right.
\nonumber\\
\!\!\!&+&\!\!\!z^2\left[\frac{\sinh \beta F_+}{F_+}\left(2J+ \frac{\Omega_{+}B_-}{B_+}+ \frac{\Omega_{-}B_+}{B_-}  \right) \right.
\nonumber\\
\!\!\!&+&\!\!\!\left.\left.
\frac{\sinh \beta F_-}{F_-}\left(2J+ \frac{\Omega_{-}B_-}{B_+}+ \frac{\Omega_{+}B_+}{B_-}  \right)\right]\right\}\;,
 \label{eq13}
 \end{eqnarray}
and $\Omega_{\pm}\!=\!(J\!\pm\!V/\sqrt{2})$. Naturally, the total uniform (staggered) magnetization are introduced
to define the order parameters for the F (AF) phase of the entire spin-electron system,
\begin{eqnarray}
m_{tot}=\frac{m_i+m_e}{1+\rho};\;\; m_{tot}^s=\frac{m_i^s+m_e^s}{1+\rho}.
\label{eq14}
\end{eqnarray}
Finally, the knowledge of the grand-canonical partition function $\Xi$~(\ref{eq6a})  can be used for a calculation of the internal energy and  the specific heat
\begin{eqnarray}
U=-\frac{\partial \ln \Xi}{\partial \beta};\;\; C=\frac{\partial U}{\partial T},
\label{eq15}
\end{eqnarray}
which allow us to study the thermal response of the spin-electron system on a decorated square lattice.

In the zero magnetic field ($h\!=\!0$), the effective field $h_{eff}$ acting within the corresponding spin-1/2 Ising model on a square lattice turns out to be zero and hence, all quantities of interest can be exactly calculated from the Onsager's exact solution for the partition function~\cite{Onsager}. Because of lack of exact solution of the spin-1/2 Ising model on a square lattice in a magnetic field we have to resort to some the-state-of-the-art numerical method. To obtain the magnetic response in the whole parametric space we have therefore used the combination of exact calculations (in the zero-field limit $h\!=\!0$) along with the state-of-the-art numerical calculation based on the Corner Transfer Matrix Renormalization Group (CTMRG) method~\cite{Nashino}.  The CTMRG technique is an accurate numerical algorithm applicable to 2D classical lattice models, which is based on the ideas of density matrix renormalization group for 1D quantum systems~\cite{White} with the  main advantage to calculate the thermodynamic behaviour in a vicinity of critical points more efficiently and accurately.
%%%%%%%%%%%%%%%%%%%%%%%%%%%%%%%%%%%
\section{Results and discussion}
\label{results}
%%%%%%%%%%%%%%%%%%%%%%%%%%%%%%%%%%%%
Before proceeding to a discussion of the most interesting results, it is necessary to note that all results are presented hereafter for the F interaction $J\!>\!0$ whose magnitude, without loss of generality, is used for normalization. In addition, the electron concentration per dimer is reduced up to the half-filled band case at most (i.e., $\rho\!\leq\!2$) due to the particle-hole symmetry. 

In order to clarify the simultaneous effect of an applied electric and magnetic fields accompanied by thermal fluctuations on  a stability of reentrance in the coupled spin-electron system, we are mainly interested in analyses of thermal magnetic phase diagrams for various model parameters.  From the previous analysis~\cite{Doria,Cenci1,Cenci3} it is known that the mutual competition in the correlated spin-electron systems can produce magnetic phase transitions with several consecutive critical points from the ordered to disordered phase or vice versa. In addition, the strength and type of interaction between localized Ising spins $J'$ has a crucial impact on the character of reentrant magnetic phase transitions with a possible existence of mixed reentrant ones~\cite{Cenci2}. Analogous effect can be observed~\cite{Cenci4} under the presence of an applied magnetic field without the direct spin-spin interaction, which stabilizes the AF ordering  (and thus generates the novel reentrance) in slightly diluted system, however, the parallel spin orientation into the magnetic field direction is expected. Last but not least, it was found that the magnetic state can be influenced by an applied electric field due to the  magnetoelectric effect and the novel reentrant magnetic transitions can occur for the another electron concentrations, even if the spin-spin interaction is switched off. In the spirit of these facts, there arise questions, which final effects will be observed when the additional spin-spin interaction is present or when the  electric and magnetic fields and temperature are simultaneously applied. Can the reentrance fully disappear or not? Does  the mixed magnetic transition persist in a vicinity of the quarter and half-filling case? 

Let us start our discussion with the particular case without the applied  magnetic field when examining solely the effect of the electric field. As was found previously, the electric field indirectly influences the subsystem of localized Ising spins and dramatically reduces the critical temperature of both ordered phases, when the system without a direct spin-spin interaction $J'$ is considered. However, the type of the F or the AF arrangement is preserved  at the quarter-filling (F) or the half-filling (AF) at the sufficiently small value of $V/J$. A detailed discussion is presented in Ref.~\cite{Cenci6}. In the case of a non-zero spin-spin interaction $J'\!\neq\!0$ the increasing electric field is responsible for the segregation of mobile electrons at bond with a dominantly non-magnetic character of the electron subsystem. Consequently, the magnetic character of the total spin-electron system is strongly determined by the magnetic character of their localized subsystem.
\begin{figure}[h!]
\begin{center}
{\includegraphics[width=0.234\textwidth,height=0.185\textheight,trim=3cm 9.2cm 4cm 9.7cm, clip]{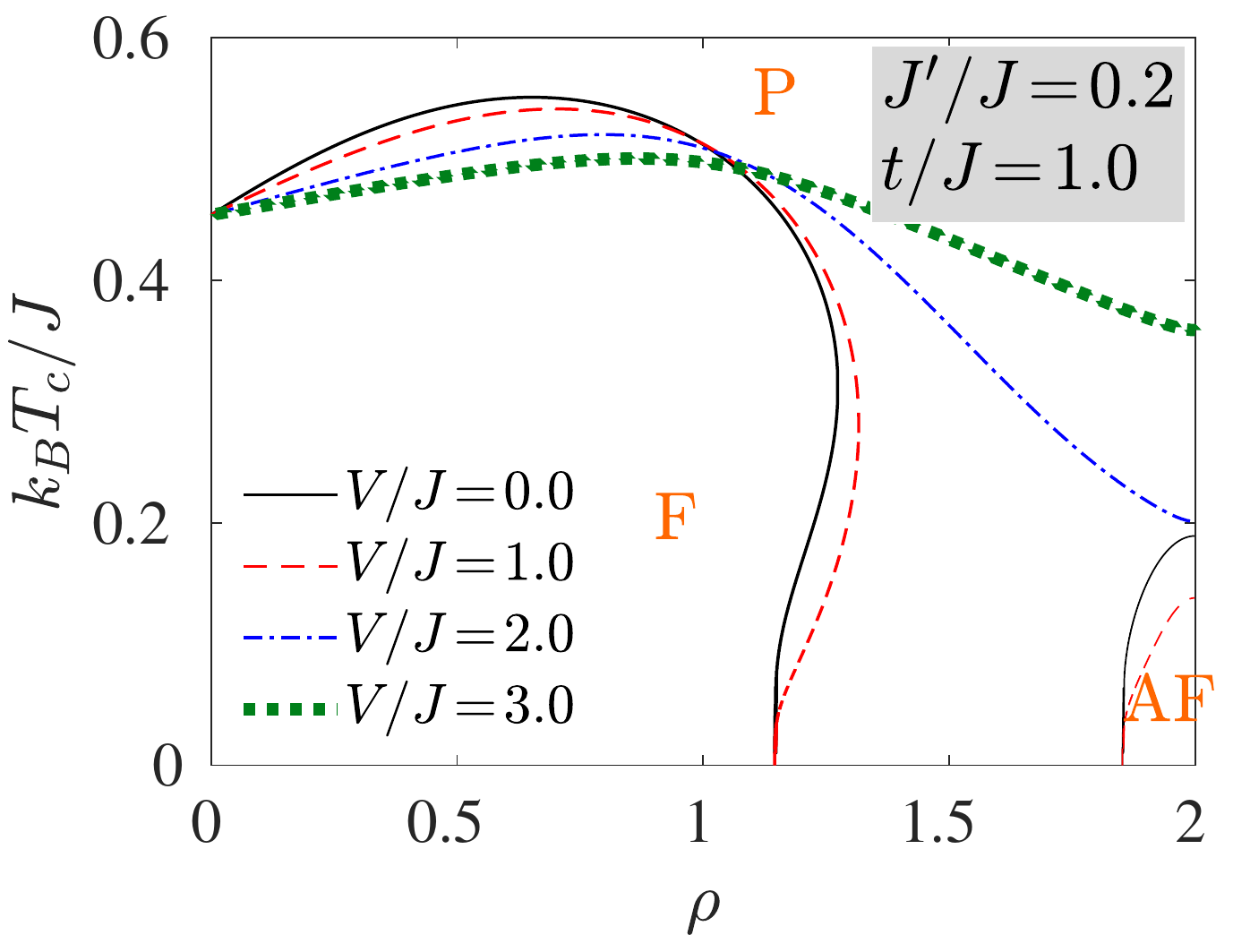}}
{\includegraphics[width=0.234\textwidth,height=0.185\textheight,trim=3cm 9.2cm 4cm 9.7cm, clip]{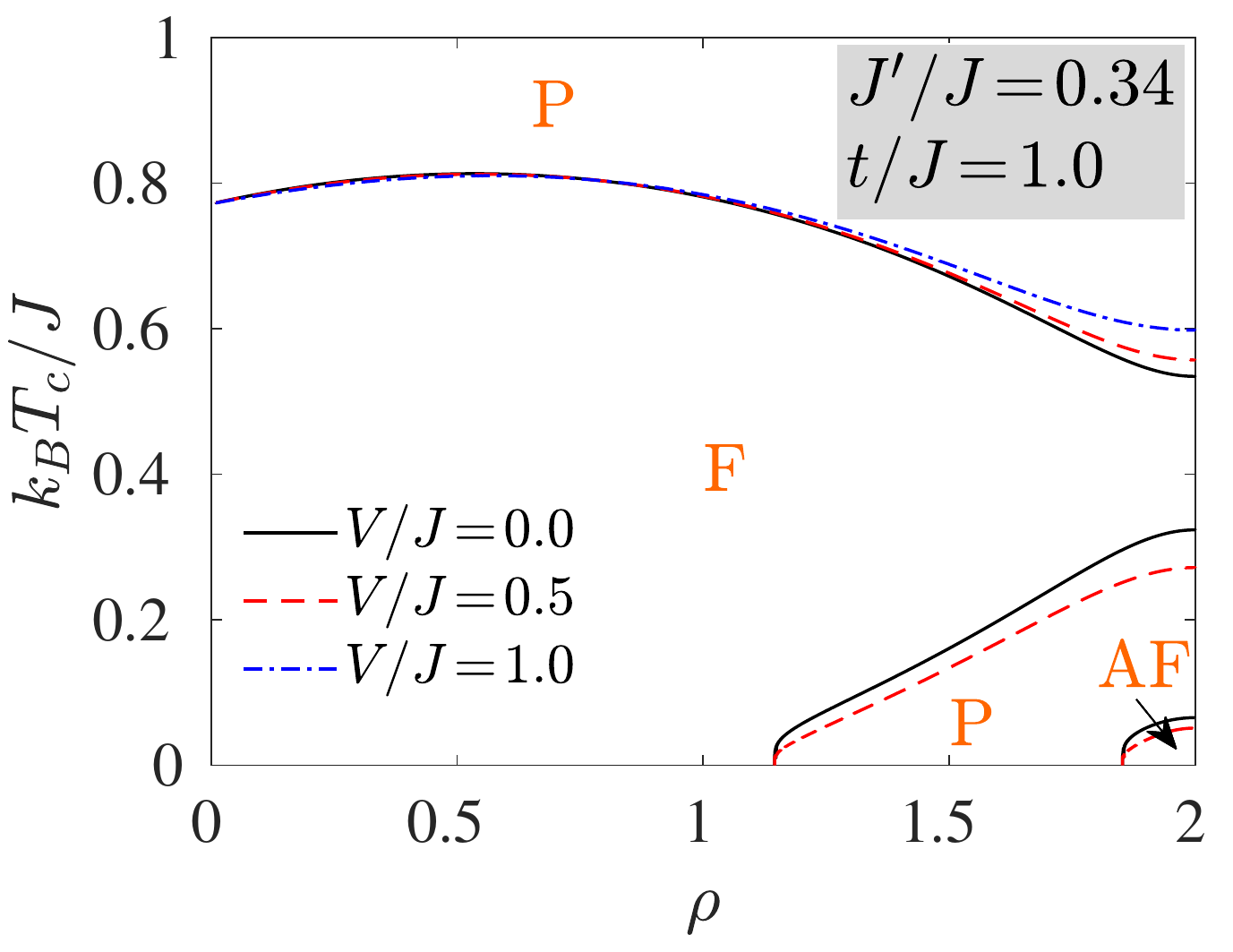}}\\
\includegraphics[width=0.234\textwidth,height=0.185\textheight,trim=3cm 9.2cm 4cm 9.7cm, clip]{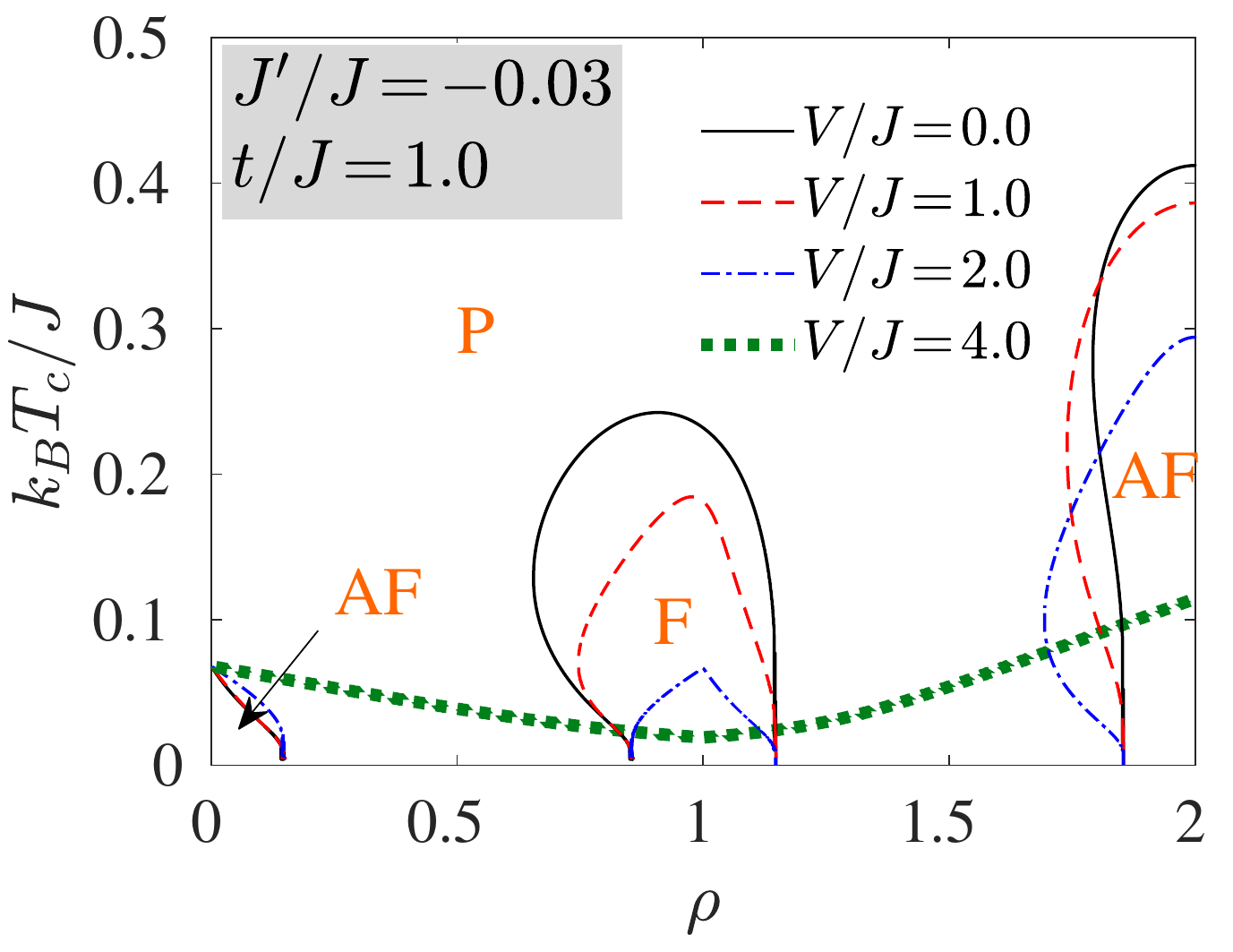}
\includegraphics[width=0.234\textwidth,height=0.185\textheight,trim=3cm 9.2cm 4cm 9.7cm, clip]{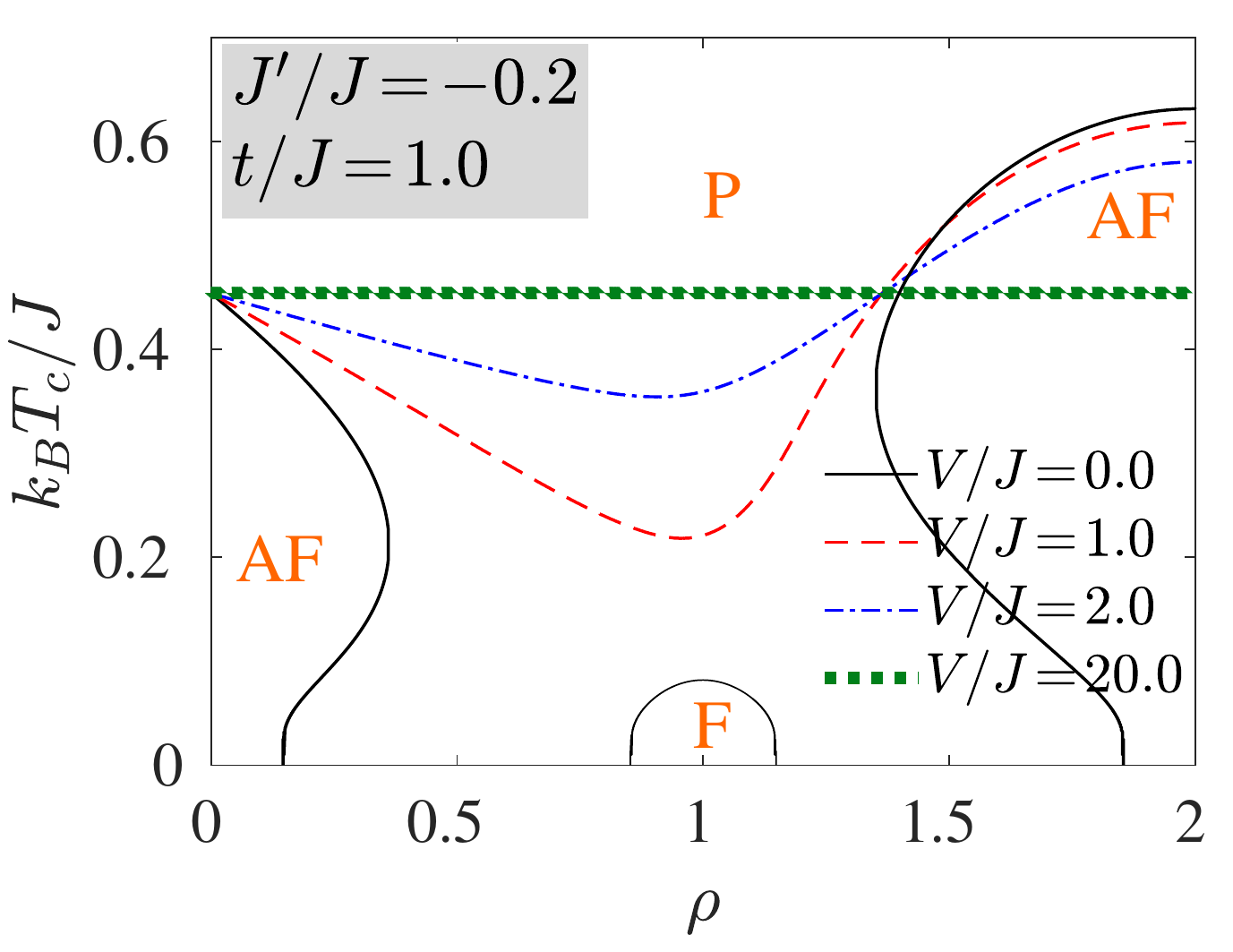}
\caption{\small  The finite-temperature phase diagrams in the $\rho\!-\!k_BT_c/J$ plane for the fixed value of the hopping parameter $t/J\!=\!1$ and different values of $J'/J$  in the zero magnetic field  ($h\!=\!0$).  Different  lines illustrate the borders between the AF-P and F-P phases for distinct values of the electrostatic potential $V/J$ (the capital letter P is used as an abbreviation for the paramagnetic phase). Exact results.}
\label{fig2}
\end{center}
\end{figure}
 In this context the applied electric field generates almost the same  effect on a thermal phase diagram as the spin-spin interaction $J'$~\cite{Cenci2}. This is nicely illustrated in Fig.~\ref{fig2}. However, a weak electric field can  generate  reentrant magnetic transitions for different concentrations $\rho$ in comparison to a zero electric-field ($V\!=\!0$) counterpart, nevertheless its consecutive increment leads to a complete destruction of the magnetic reentrance. This similar effect of an applied electric field  $V$ and a spin-spin interaction $J'$ seems to be very interesting consequence, because  reentrant magnetic phase transitions can be more easily tuned by an applied electric field rather than the synthesis of magnetic materials with different atoms or pressure-induced change of the coupling ratio $J'/J$. 
 
Another fascinating observation is that there exists a critical value of the electrostatic potential $V/J$, above which the critical temperature is identical for any electron concentration $\rho$, see the right-down panel of Fig.~\ref{fig2}. It was found that this special value of the critical temperature, is determined by a relative simple relation $k_BT_c/J\!=\!2J'/J\ln(1+\sqrt{2})$, which depends only on a relative strength of the further-neighbour spin-spin interaction $J'/J$ and the underlying magnetic lattice. This interesting phenomenon opens a new possibility to use the quasi-2D spin-electron systems with an exactly defined order-disorder critical  temperature, independent on the electron doping. 

\begin{figure}[b!]
\begin{center}
{\includegraphics[width=0.33\textwidth,height=0.22\textheight,trim=3cm 9.2cm 4.cm 9.5cm, clip]{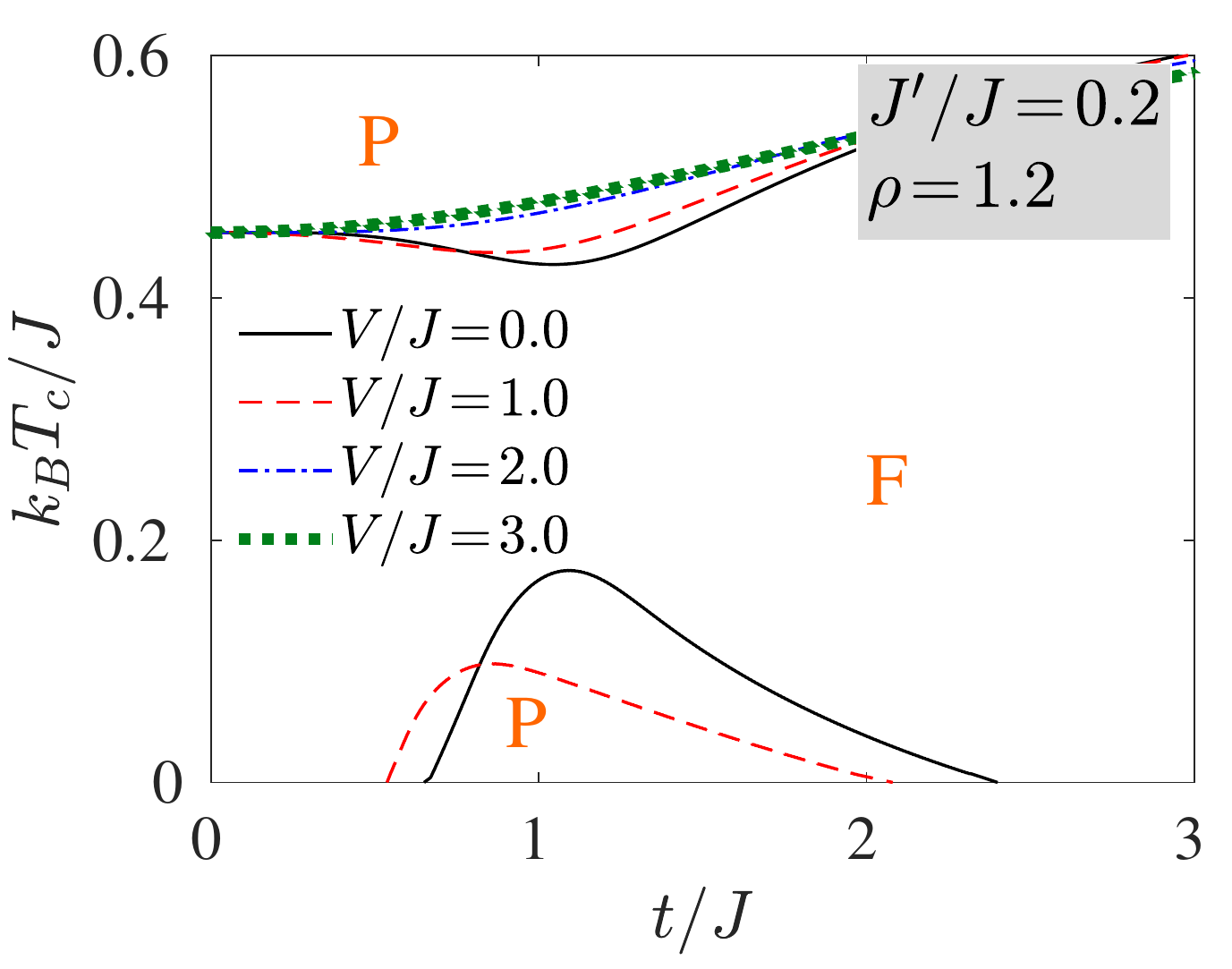}}
\includegraphics[width=0.33\textwidth,height=0.22\textheight,trim=3cm 9.2cm 4.cm 9.5cm, clip]{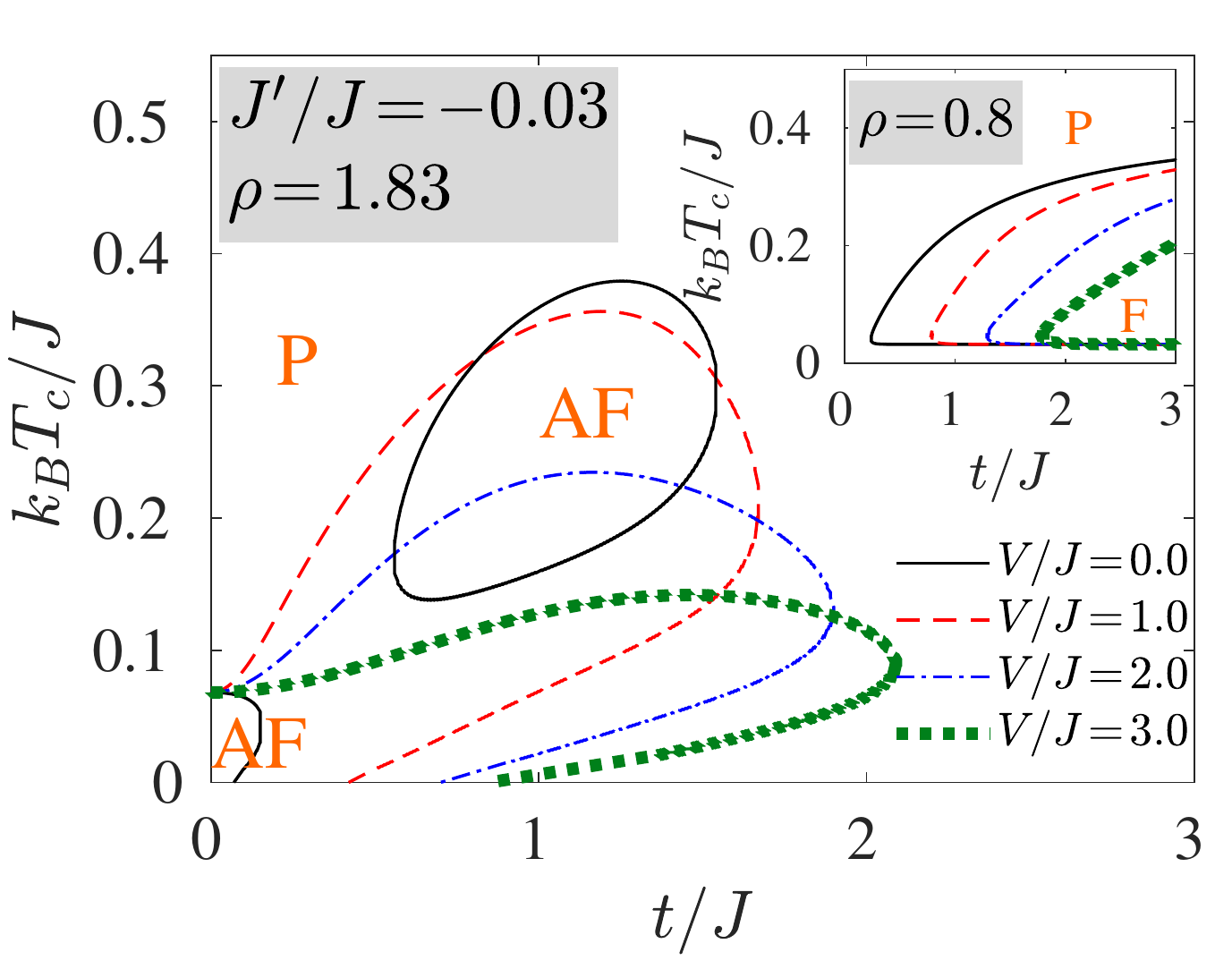}
\includegraphics[width=0.33\textwidth,height=0.22\textheight,trim=3cm 9.2cm 4.cm 9.5cm, clip]{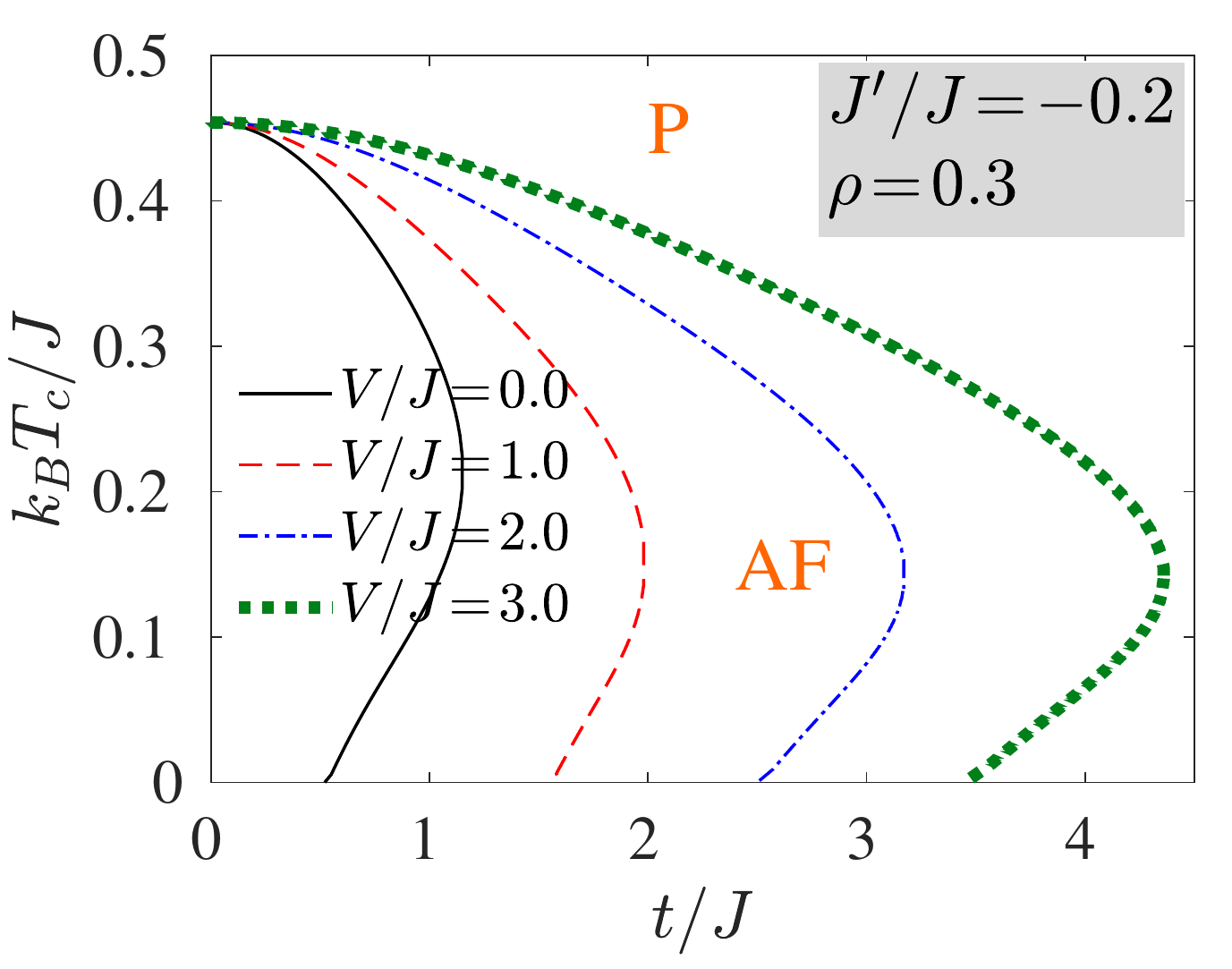}
\caption{\small  The finite-temperature phase diagrams in the $t/J\!-\!k_BT_c/J$ plane for different values of $J'/J$ and electron concentration $\rho$ in the zero magnetic field  ($h\!=\!0$).  Different  lines illustrate the borders between the AF-P and F-P phases for distinct values of the electric field $V/J$. Exact results.}
\label{fig3}
\end{center}
\end{figure}

As we known from previous analysis~\cite{Cenci1,Doria}, the nonnegligible effect on a stability of reentrant magnetic transitions originates from the hopping process of mobile electrons. It is reasonable to suppose that under presence of the electric field, which directly influences the mobile electrons, this effect could be even more pronounced. For this reason, let us  analyse in detail the behavior of the critical temperature as a function of the electron hopping. Typical plots are presented in Fig.~\ref{fig3}.
 Indeed, the applied electric field can dramatically change the critical temperature and moreover, the competition between the interaction parameters $V$, $J$ and $J'$ produces a reentrant behavior at novel values of the hopping amplitude $t$. The most obvious changes occur in the AF phase ($J'/J\!=\!-0.2$) with a low concentration of the mobile electrons ($\rho\!=\!0.3$), where the reentrance is shifted from the lower value of the hopping amplitude ($t/J\!\approx\!1$) to the higher ones ($t/J\!\approx\!4$).

\begin{figure}[b!]
\begin{center}
{\includegraphics[width=0.234\textwidth,height=0.185\textheight,trim=3cm 9.4cm 4cm 9.5cm, clip]{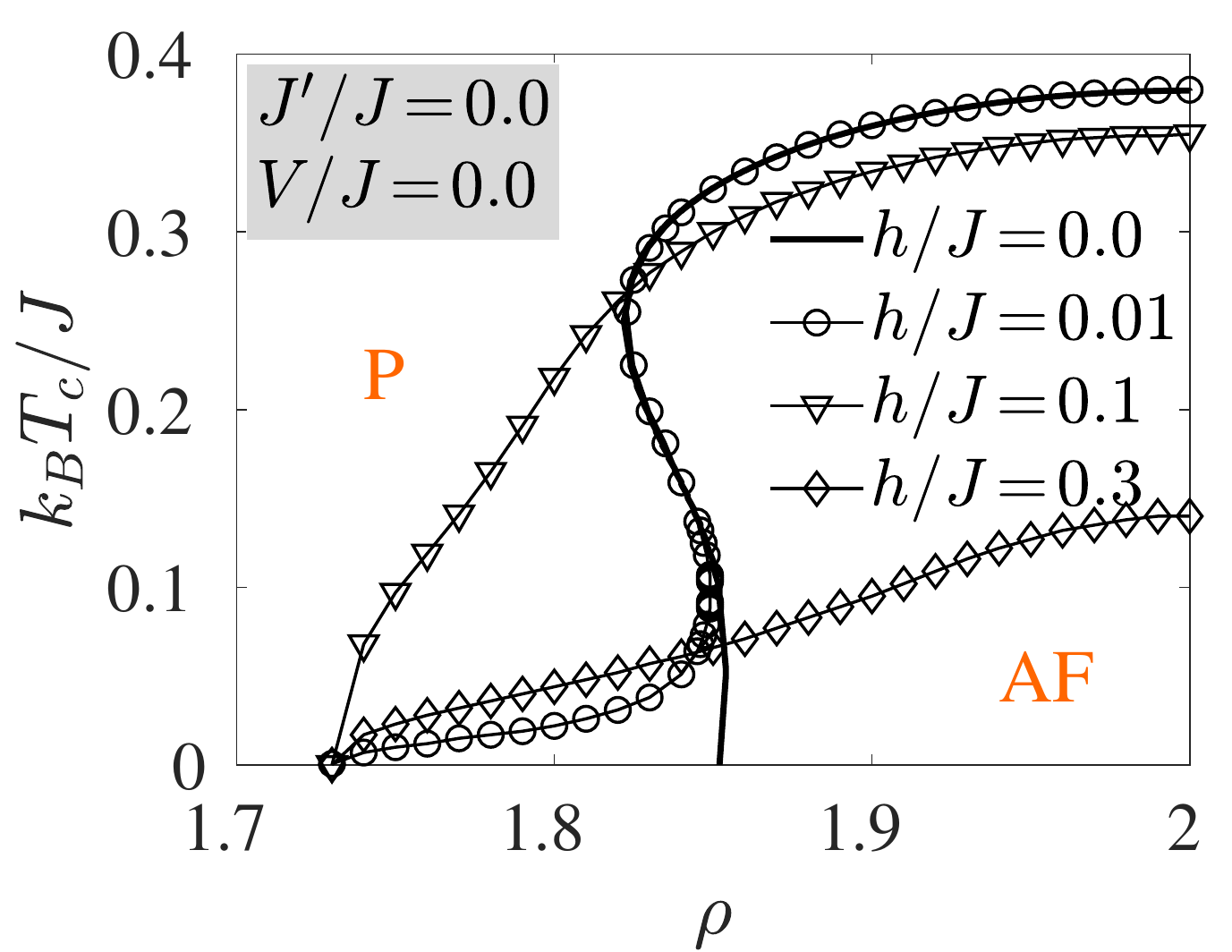}}
{\includegraphics[width=0.234\textwidth,height=0.185\textheight,trim=3cm 9.4cm 4cm 9.5cm, clip]{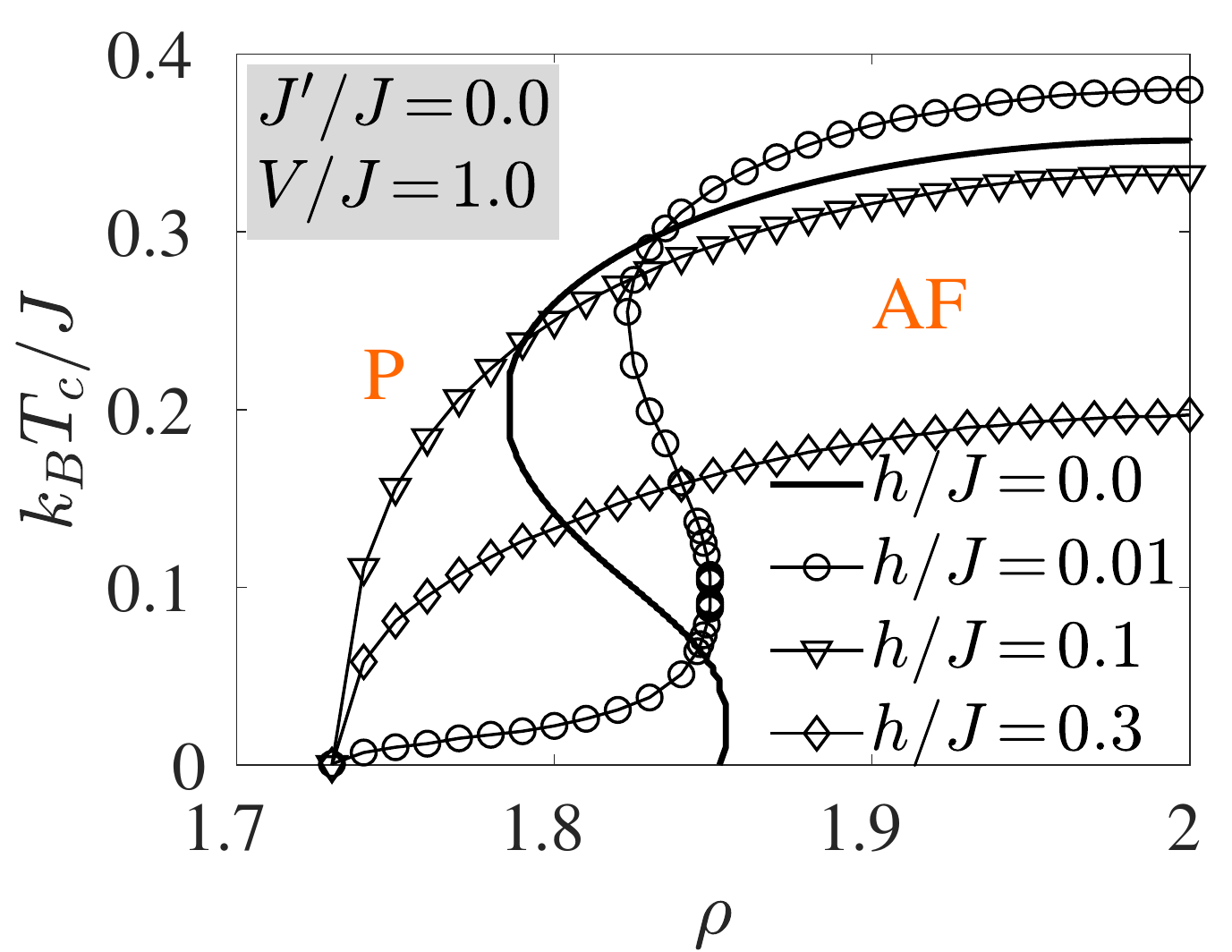}}\\
{\includegraphics[width=0.234\textwidth,height=0.185\textheight,trim=3cm 9.4cm 4cm 9.5cm, clip]{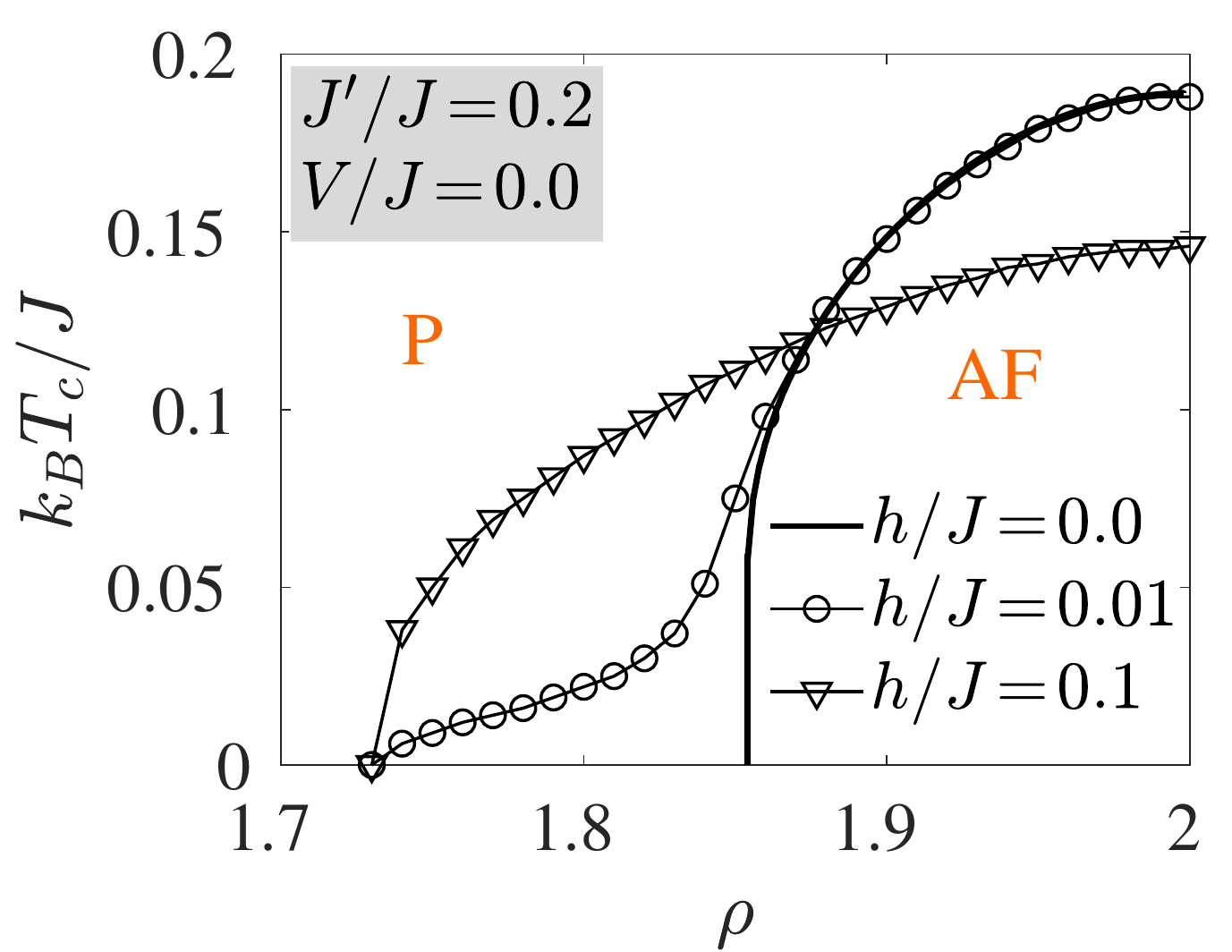}}
{\includegraphics[width=0.234\textwidth,height=0.185\textheight,trim=3cm 9.4cm 4cm 9.5cm, clip]{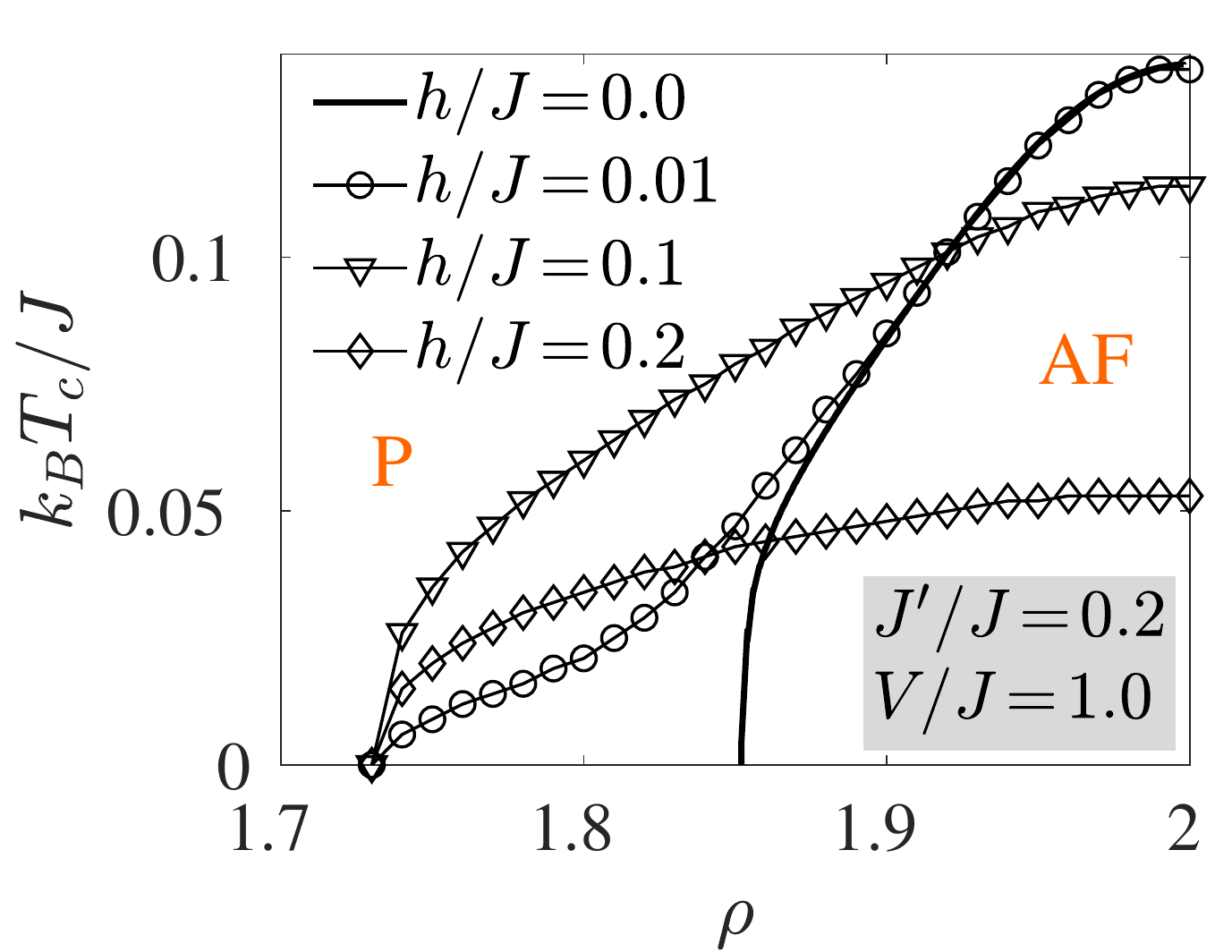}}
\caption{\small The finite-temperature phase diagrams  in the $\rho\!-\!k_BT_c/J$ plane for various values of the model parameters and selected values of the magnetic field $h/J$ given in the legend. Exact results (solid lines without symbols) and CTMRG results (lines with symbols). }
\label{fig4}
\end{center}
\end{figure}

\begin{figure}[t!]
\begin{center}
{\includegraphics[width=0.234\textwidth,height=0.185\textheight,trim=3cm 9.4cm 4cm 9.5cm, clip]{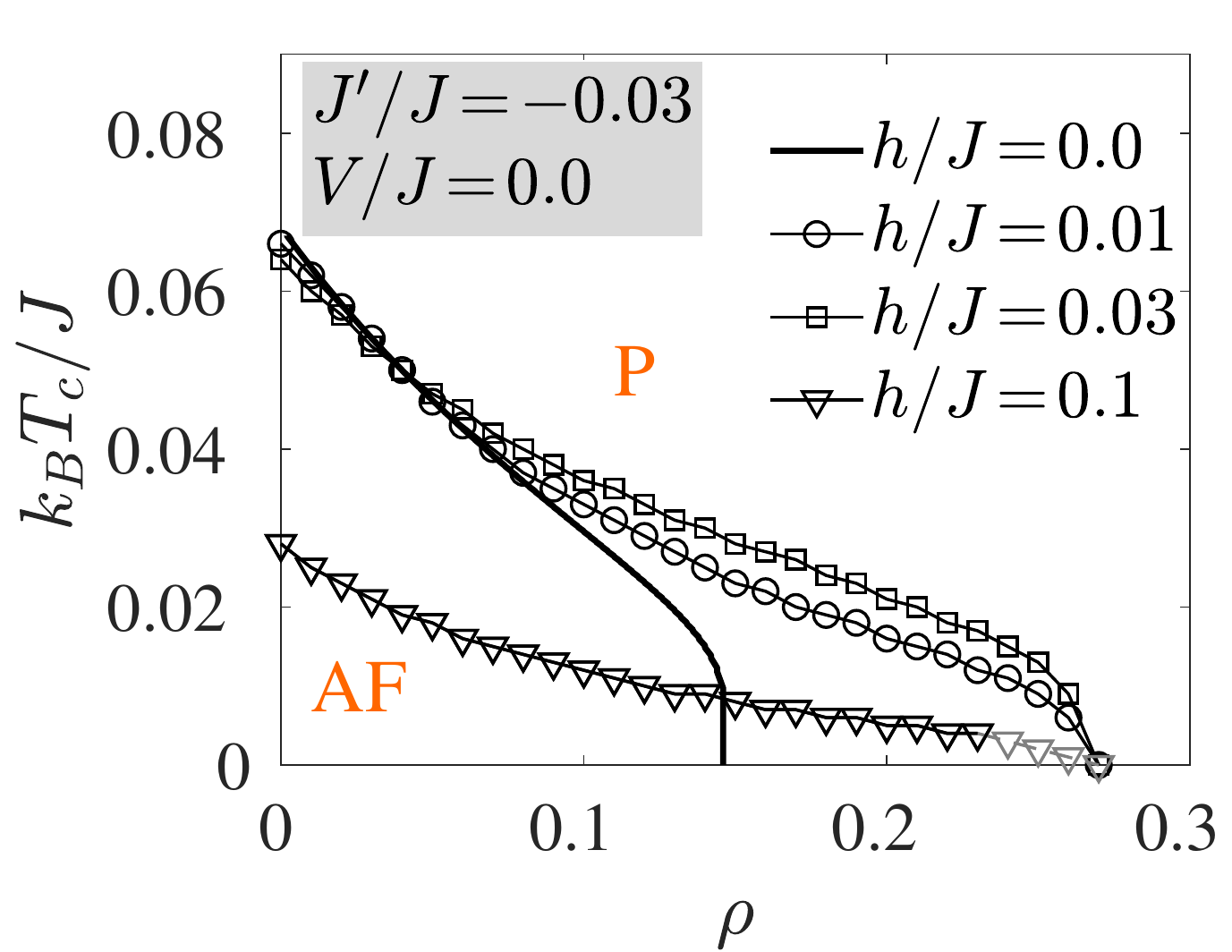}}
{\includegraphics[width=0.234\textwidth,height=0.185\textheight,trim=3cm 9.4cm 4cm 9.5cm, clip]{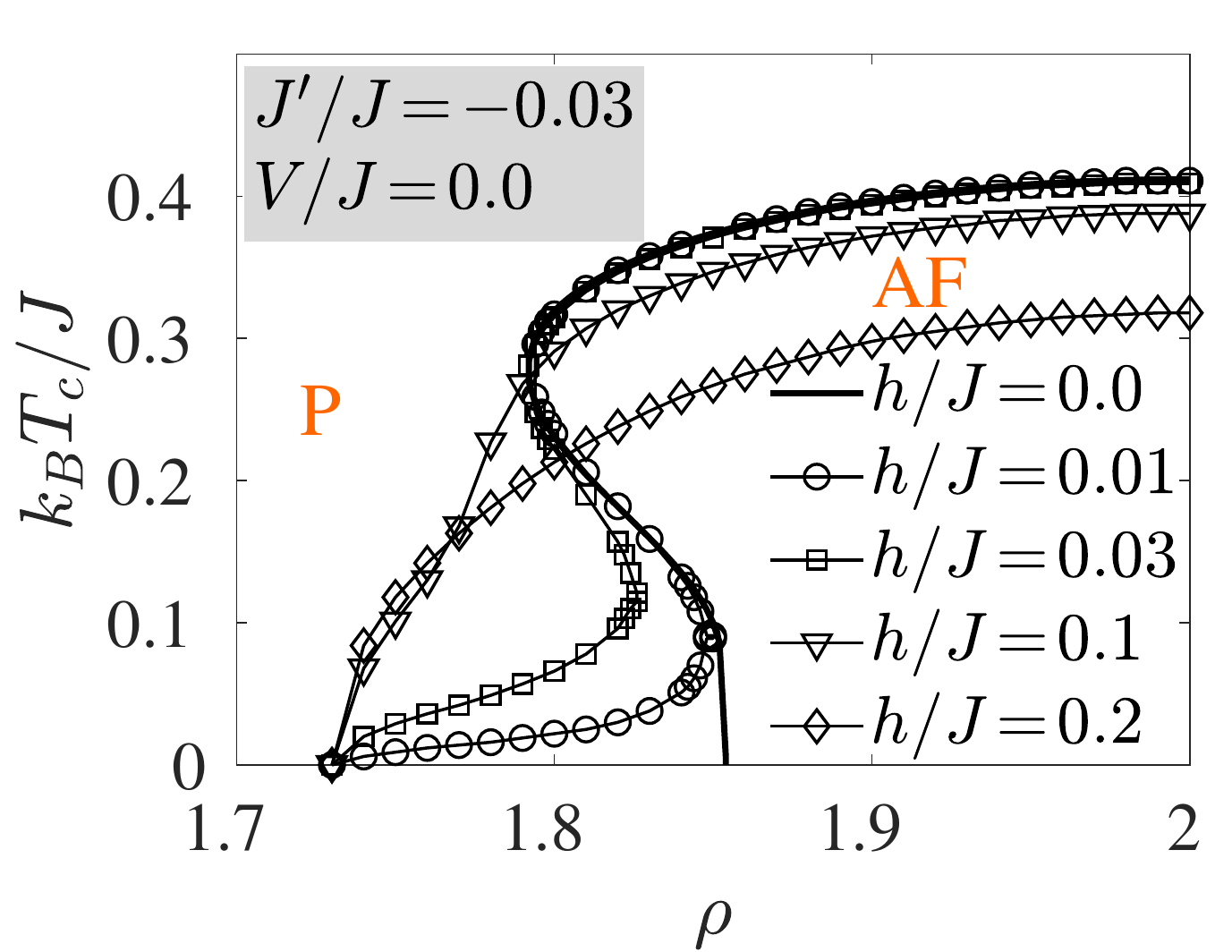}}
{\includegraphics[width=0.234\textwidth,height=0.185\textheight,trim=3cm 9.4cm 4cm 9.5cm, clip]{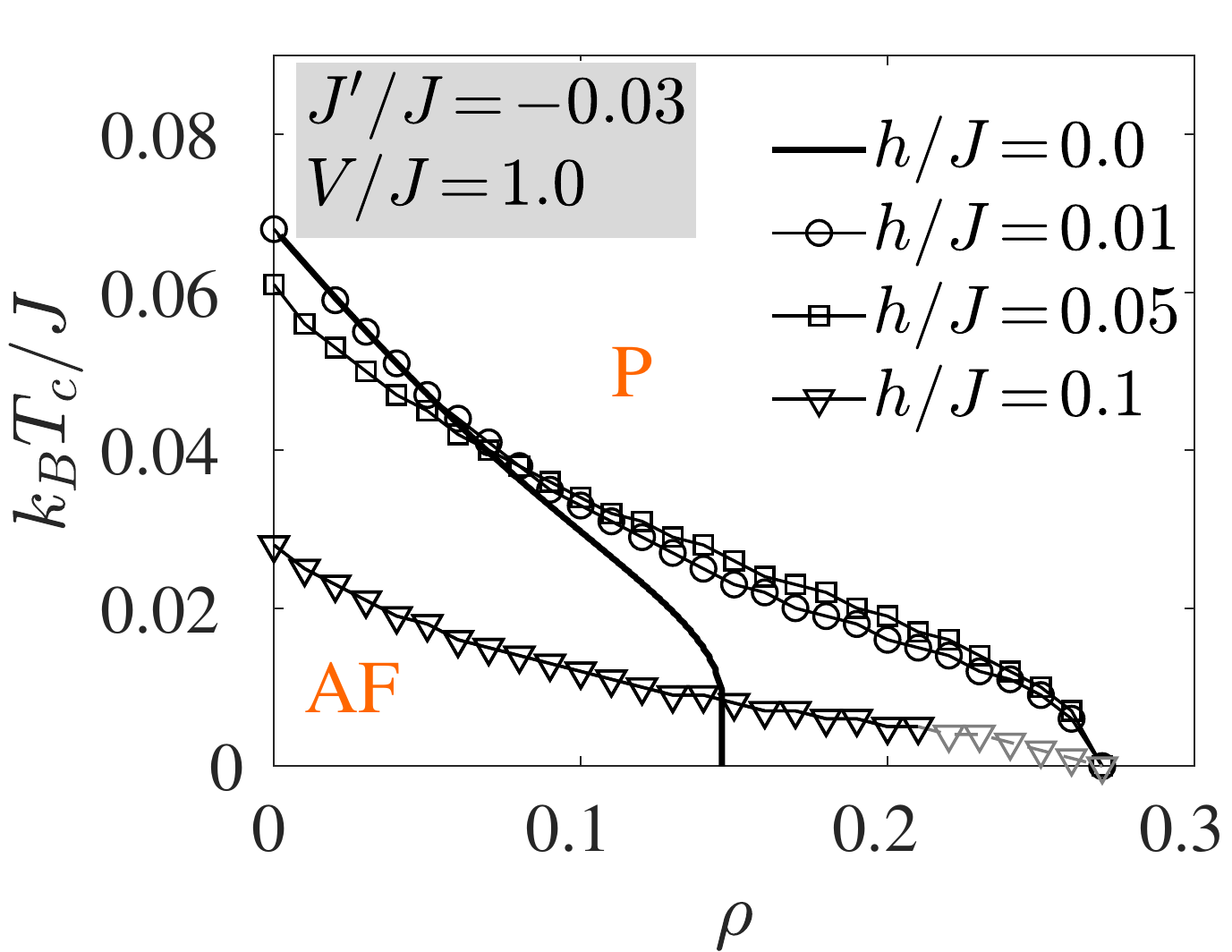}}
{\includegraphics[width=0.234\textwidth,height=0.185\textheight,trim=3cm 9.4cm 4cm 9.5cm, clip]{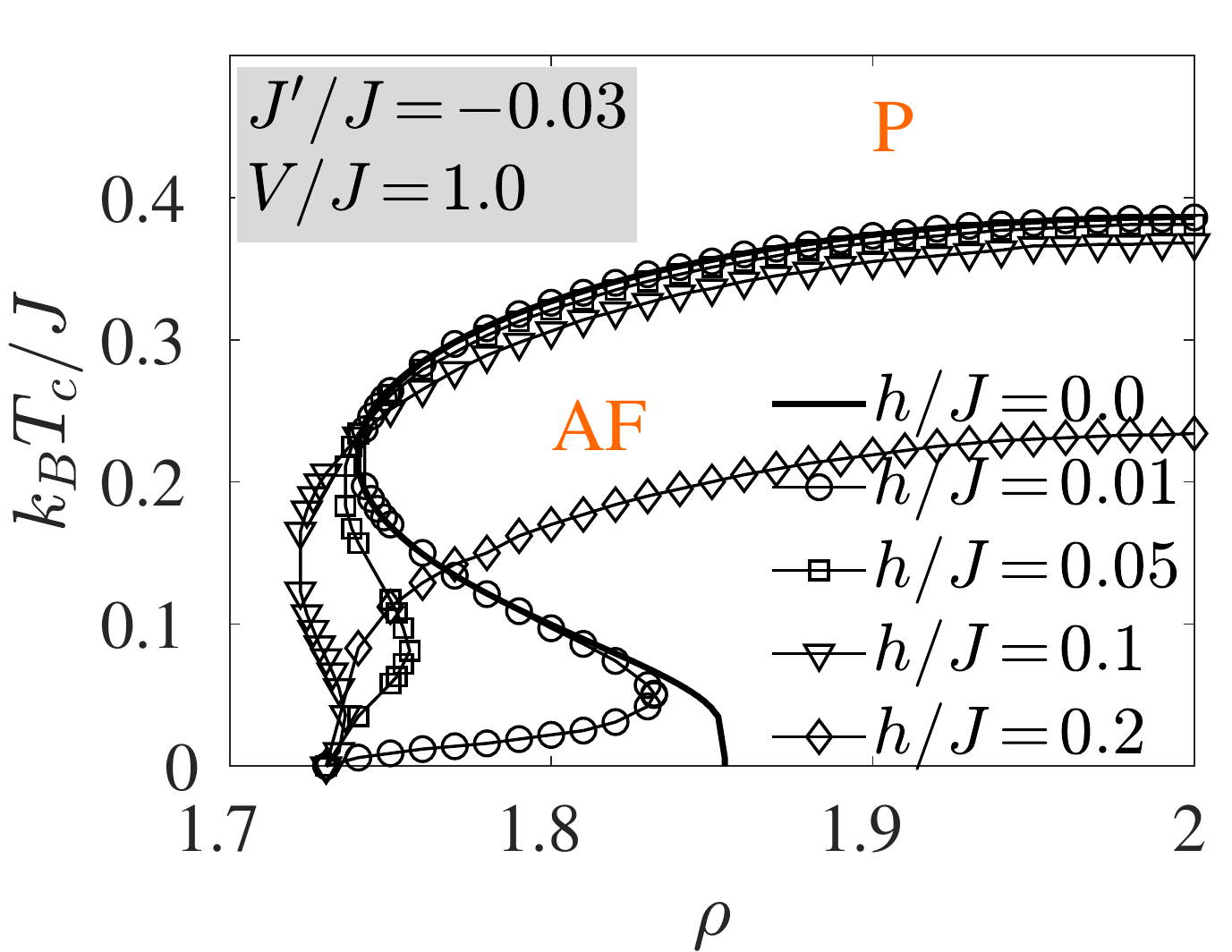}}\\
{\includegraphics[width=0.234\textwidth,height=0.185\textheight,trim=3cm 9.4cm 4cm 9.5cm, clip]{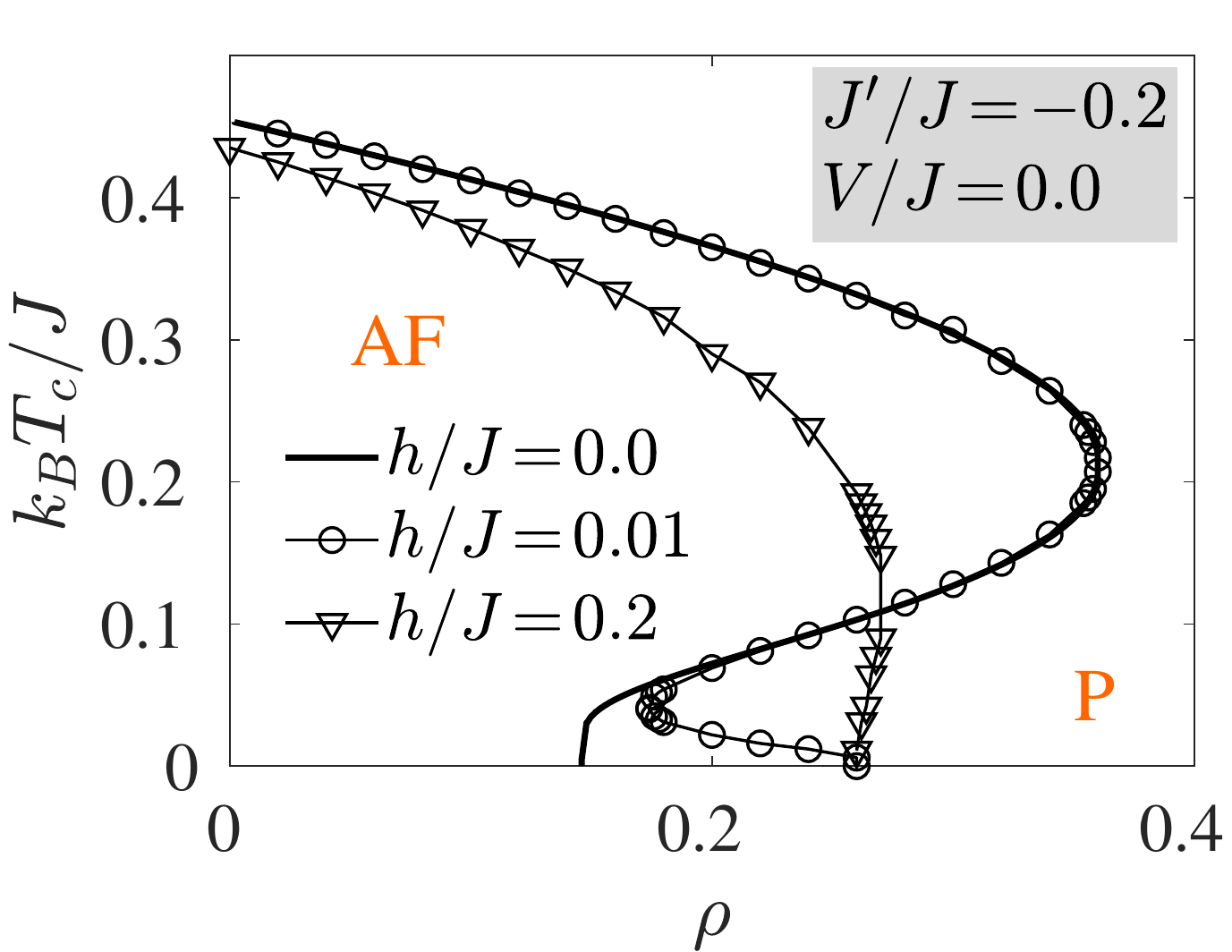}}
{\includegraphics[width=0.234\textwidth,height=0.185\textheight,trim=3cm 9.4cm 4cm 9.5cm, clip]{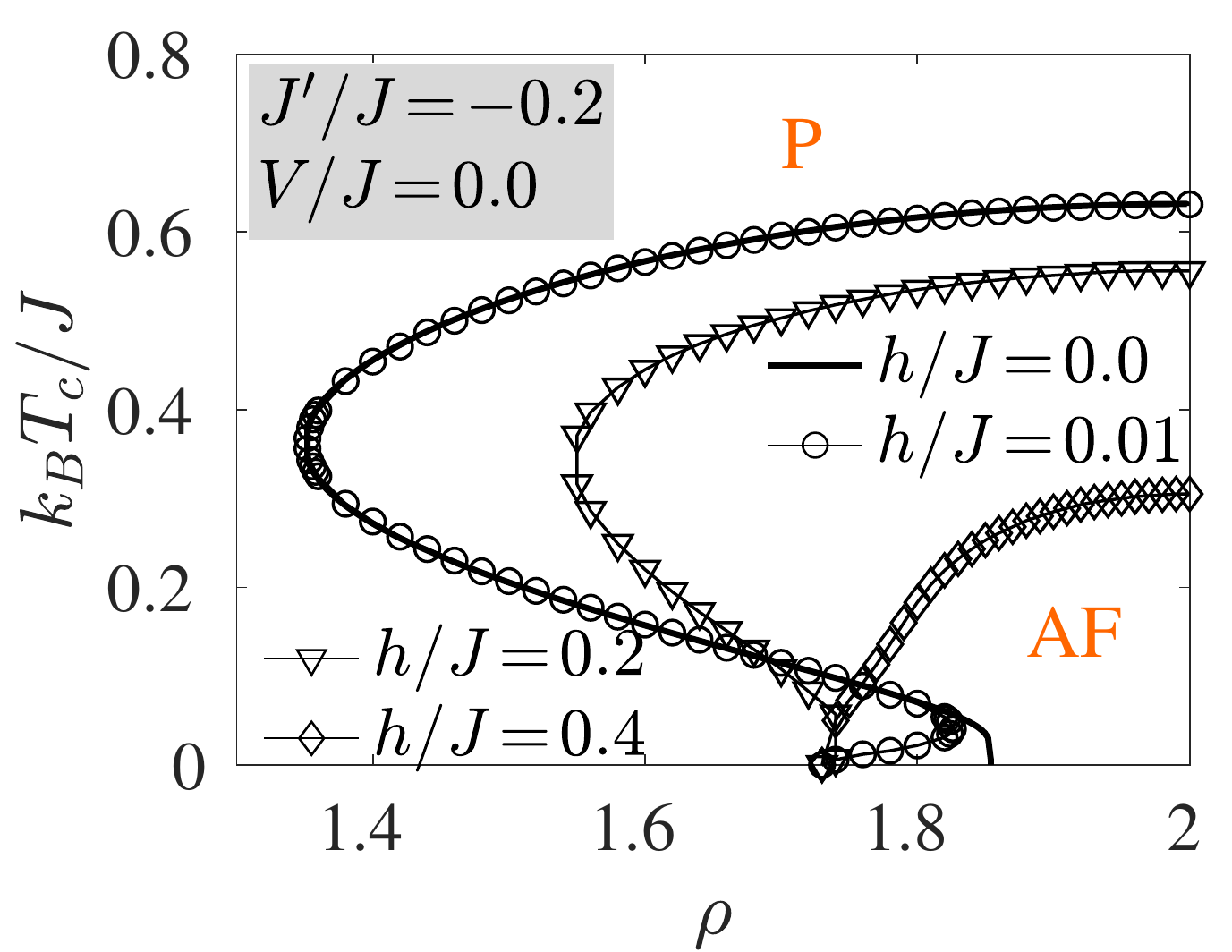}}
\\
{\includegraphics[width=0.234\textwidth,height=0.185\textheight,trim=3cm 9.4cm 4cm 9.5cm, clip]{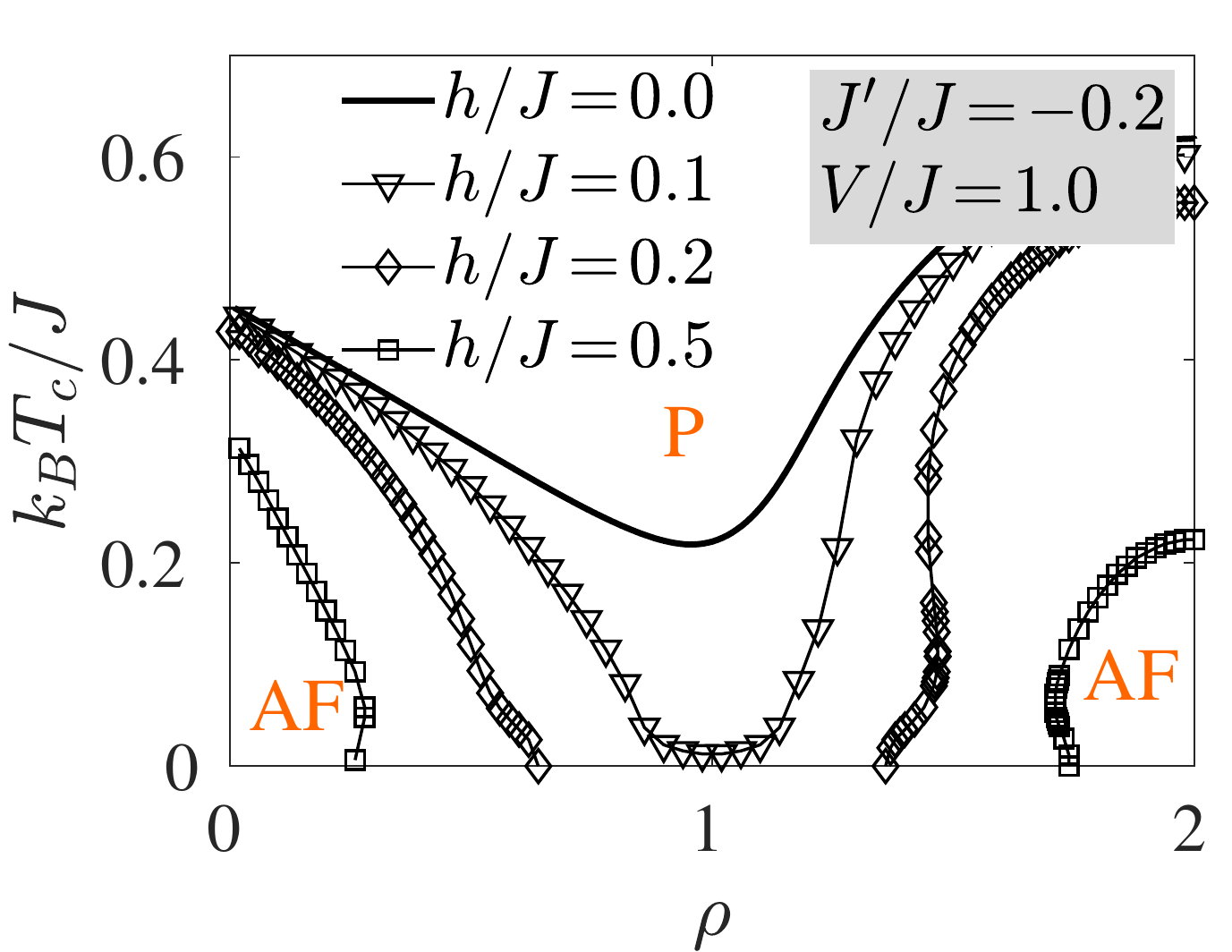}}
{\includegraphics[width=0.234\textwidth,height=0.185\textheight,trim=3cm 9.4cm 4cm 9.5cm, clip]{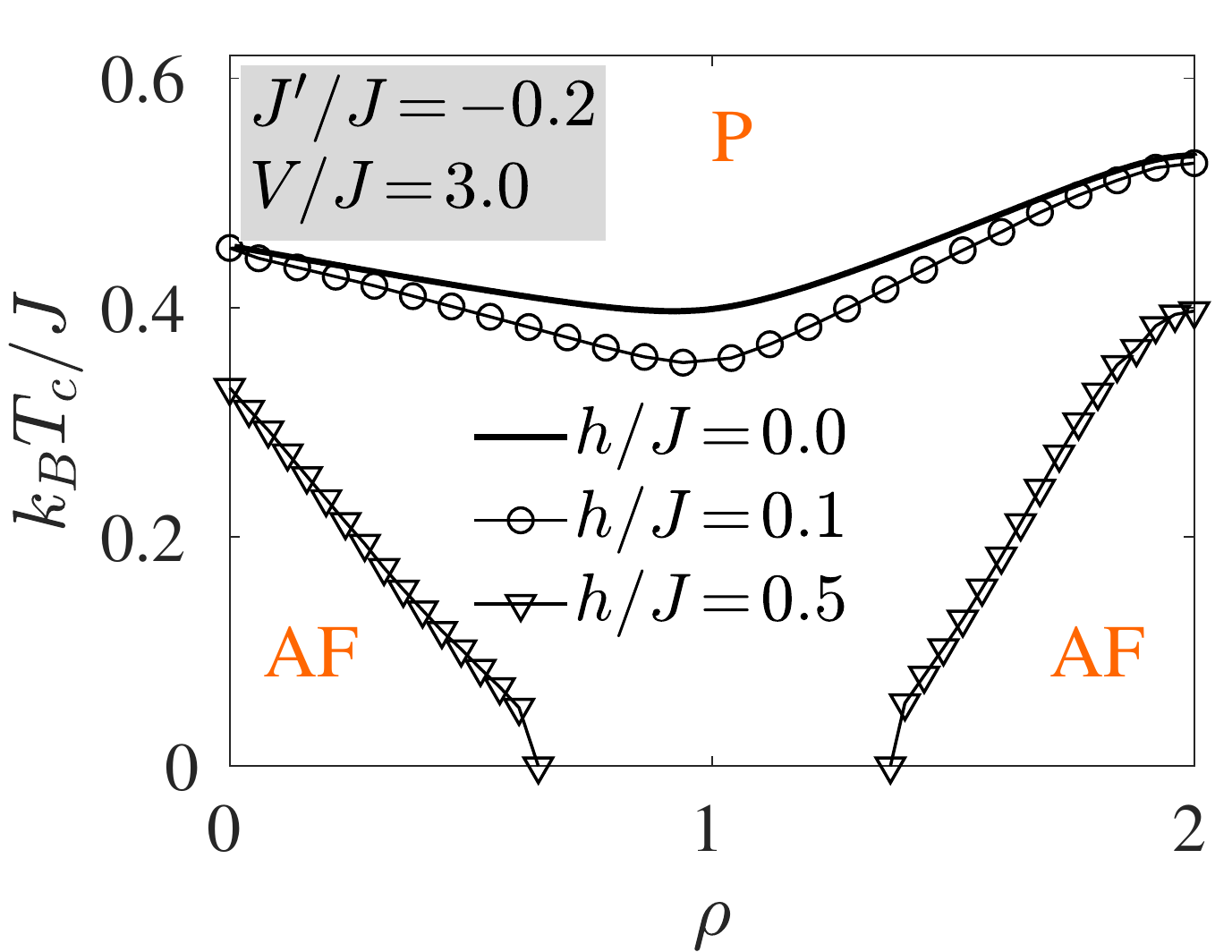}}
\caption{\small  The finite-temperature phase diagrams  in the $\rho\!-\!k_BT_c/J$ plane for various values of the model parameters and selected values of the magnetic field $h/J$ given in the legend. Exact results (solid lines without symbols) and CTMRG results (lines with symbols). }
\label{fig5}
\end{center}
\end{figure}

Next, let us apply the non-zero external magnetic field $h$, which acts on the localized Ising spins as well as the mobile electrons. In agreement with our expectations, the magnetic field causes a reorientation of almost all randomly oriented spins into its direction and it also gradually suppressed the AF arrangement. However, the applied magnetic field may also transform a randomly oriented spins into the AF type of ordering and thus, generate novel reentrant transitions with three consecutive critical points instead of two consecutive critical points in a vicinity of critical concentrations $\rho_c$ detected in zero-magnetic field ($h\!=\!0$) at the electron concentrations close to a half filling, see Figs.~\ref{fig4} and~\ref{fig5}. 
The magnetic field may thus strikingly cause emergence of the spontaneous AF long-range order at slightly lower electron densities $\rho\!=\!1.736$ in comparison with the critical concentration $\rho^{AF}_c\!=\!1.854$ detected in zero magnetic field. It is quite interesting that reentrant region is identical for an arbitrary $J'/J$ and appropriate $V/J$ for sufficiently small $h/J\!\sim\!0.01$ (appropriate $V/J$ means the region where the ground state for $\rho\!=\!1$ ($\rho\!=\!2$) corresponds to the F (AF) ordering). It could be anticipated that this unconventional behavior originates from the fact that a slightly diluted electron system involves two kinds of bonds: the majority of bonds ($p$) are occupied by two mobile electrons per dimer that mediate the AF coupling and the smaller fraction of bonds $r$ ($r\!<\!p$) involves just  one mobile electron per dimer that mediates the  F coupling. Since we have a slightly over-doped electron system, a few AF bonds are distributed among the F ones. 
Then, the applied magnetic field requires reorientation of magnetic moments into its direction, which results to the immediate increment of the staggered magnetization $m^s_{tot}$. In addition, the non-zero magnetic field $h$ leads to the splitting of doubly occupied electron states, which dramatically changes their non-magnetic character. In general, the simultaneous influence of the electric and magnetic fields is responsible for a declination of a critical temperature, but it does not have a dramatic influence on the existence of reentrant transitions for systems with a non-negative spin-spin interaction. This fact can be seen in Fig.~\ref{fig4}, which presents a few typical behaviors for selected model parameters. For the AF spin-spin interaction the situation is slightly different. If the AF interaction is relatively small e.g. $J'/J\!=\!-0.03$ as presented in Fig.~\ref{fig5}, thermal phase diagrams undergo similar changes as reported for the F spin-spin interaction. In addition, the modification of the phase boundaries at $\rho\!\to\!0$ under the influence of an electric field $V$ and magnetic field $h$ has a similar scenario (the gray symbols in Fig.~\ref{fig5} correspond to extrapolated values, which are unobtainable due to very low temperature). The magnetic field generally suppressed the critical temperature of the AF phase, while the lower critical concentration $\rho_c^{AFL}$ shifts from $\rho\!=\!0.146$ to $\rho\!=\!0.262$ in presence of the magnetic field irrespective of the external electric field. On the other hand, the novel reentrant transitions  occur for the sufficiently large spin-spin interaction $|J'|/J$ and the electric field $V/J$.
Fig.~\ref{fig5} illustrates one typical example for $J'/J\!=\!-0.2$ and $V/J\!=\!1$, where two reentrant regions of the AF-P-AF-P type at $\rho\!\approx\!1.5$ or of the P-AF-P type for $\rho\!<\!0.4$ are present at $h/J\!=\!0.2$ and $h/J\!=\!0.5$, respectively (the symbol P denotes a disordered paramagnetic phase). In addition, the mutual influence of the electric and magnetic fields on the coupled spin-electron system has a dramatic impact on the modification of a critical concentration, as it is evident from Fig.~\ref{fig5}. 
\begin{figure}[t!]
\begin{center}
{\includegraphics[width=0.234\textwidth,height=0.185\textheight,trim=3.cm 9.4cm 4cm 9.5cm, clip]{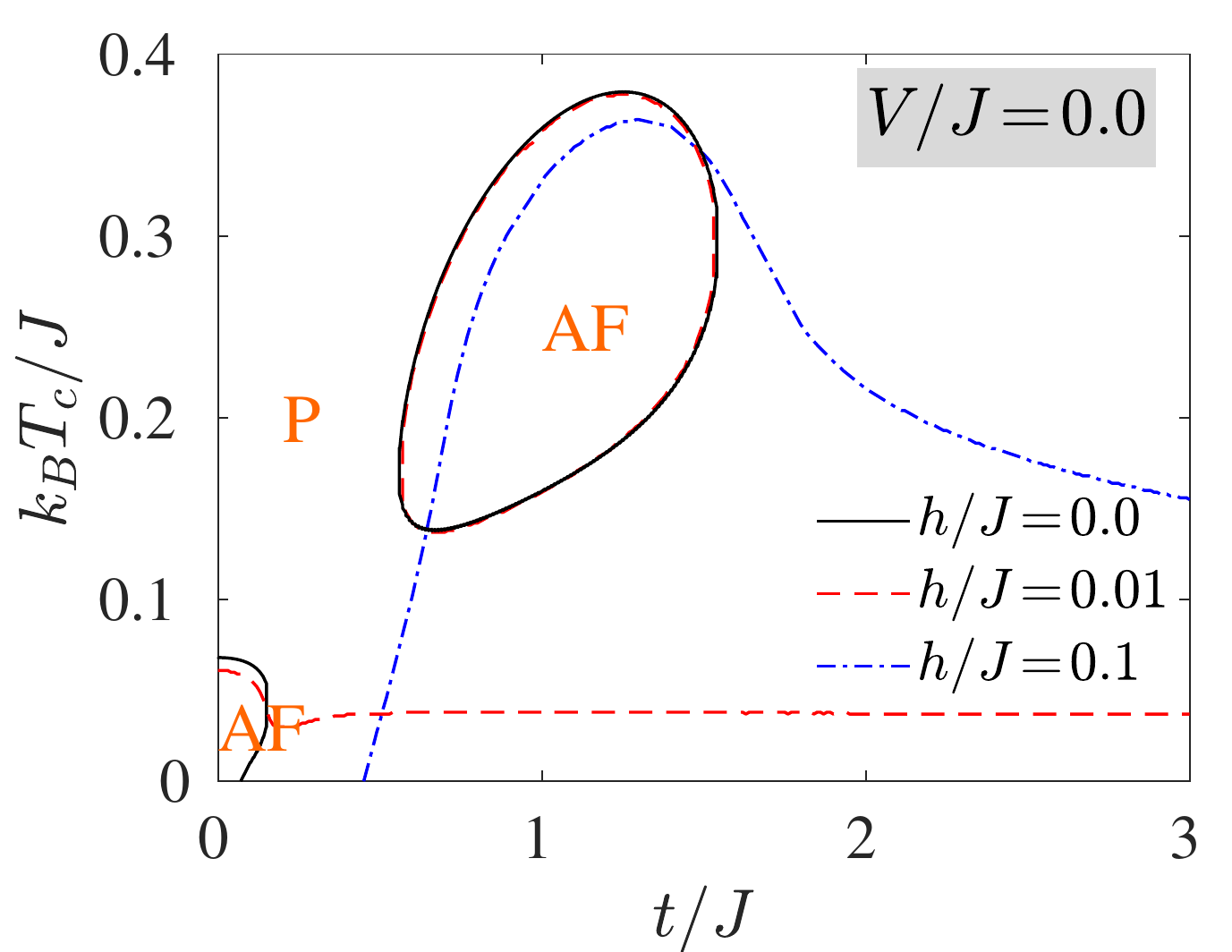}}
{\includegraphics[width=0.234\textwidth,height=0.185\textheight,trim=3.cm 9.4cm 4cm 9.5cm, clip]{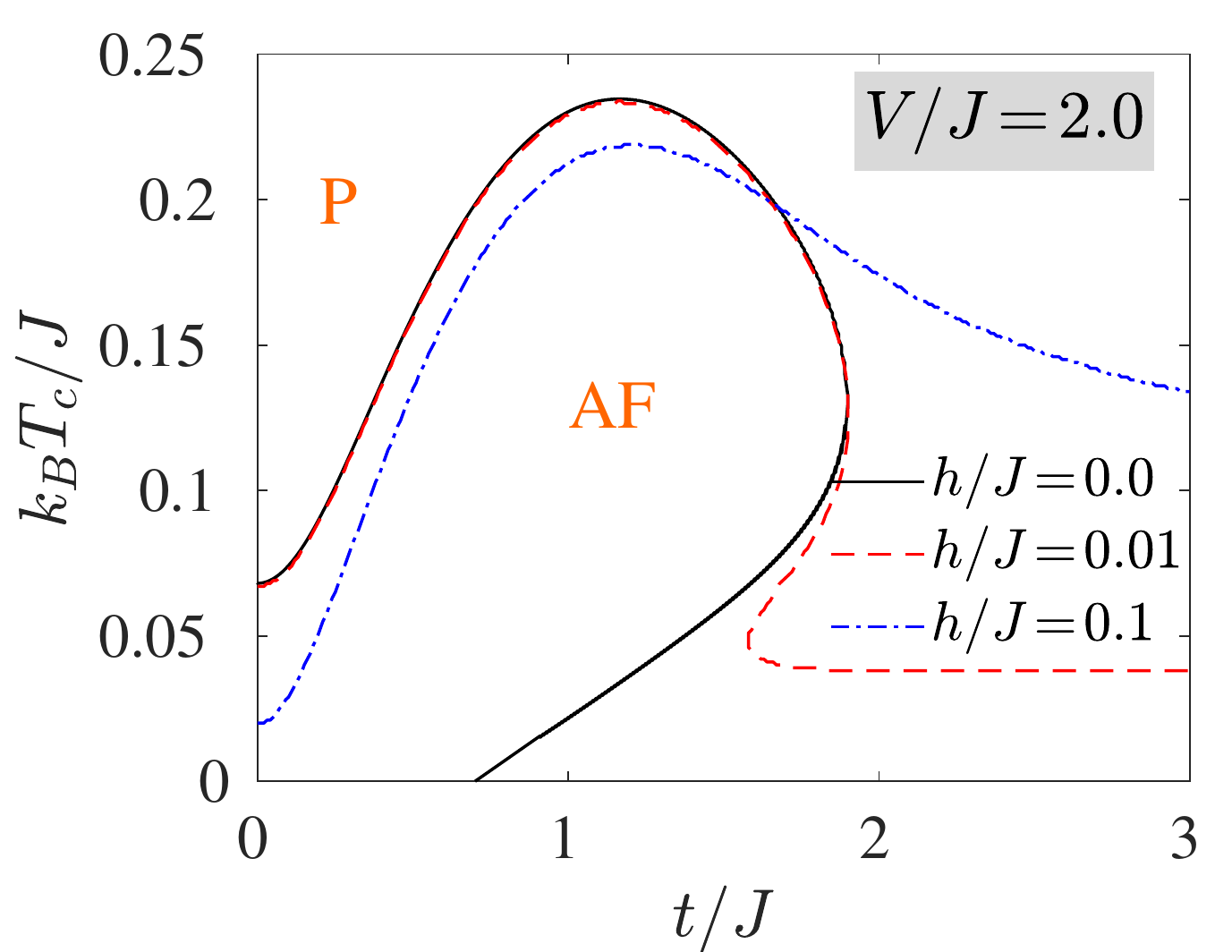}}
\caption{\small  The finite-temperature phase diagrams in the $t/J\!-\!k_BT_c/J$ plane for fixed values of the electron concentration $\rho\!=\!1.83$ and the spin-spin interaction $J'/J\!=\!-0.03$. The influence of an applied magnetic field (given in the legend) is illustrated for two different values of the electrostatic potential $V/J\!=\!0$ and $V/J\!=\!2$.}
\label{fig6}
\end{center}
\end{figure}
It is also worthy to examine, how the magnetic and electric fields separately or simultaneously shift the region of reentrant transitions depending on a relative size of the hopping amplitude. One interesting example  is visualized in Fig.~\ref{fig6} for $\rho\!=\!1.83$ with the most pronounced results.  At first, the applied magnetic field reduces the existence of the ordered phase at low hopping amplitudes $t/J\!\to\! 0$, but stabilizes its existence at higher ones.  One can also detect that  the magnetic field leads to the decrease of the critical temperature for $t/J\!\lesssim\!1$, while above $t/J\!\gtrsim\!1$ an opposite trend is observed.
 Also in this case the most interesting result from these observations lies in a new simpler alternative mechanism how to vary the magnetic reentrant transitions in coupled spin-electron systems in two dimensions. 
It is interesting to note that the involved correlation effects can stabilize various types of the AF ordering in weakly doped spin-electron system, if the AF spin-spin interaction is assumed.
\begin{figure}[t!]
\begin{center}
\includegraphics[width=0.254\textwidth,height=0.175\textheight,trim=3.cm 9.3cm 3.cm 9.8cm, clip]{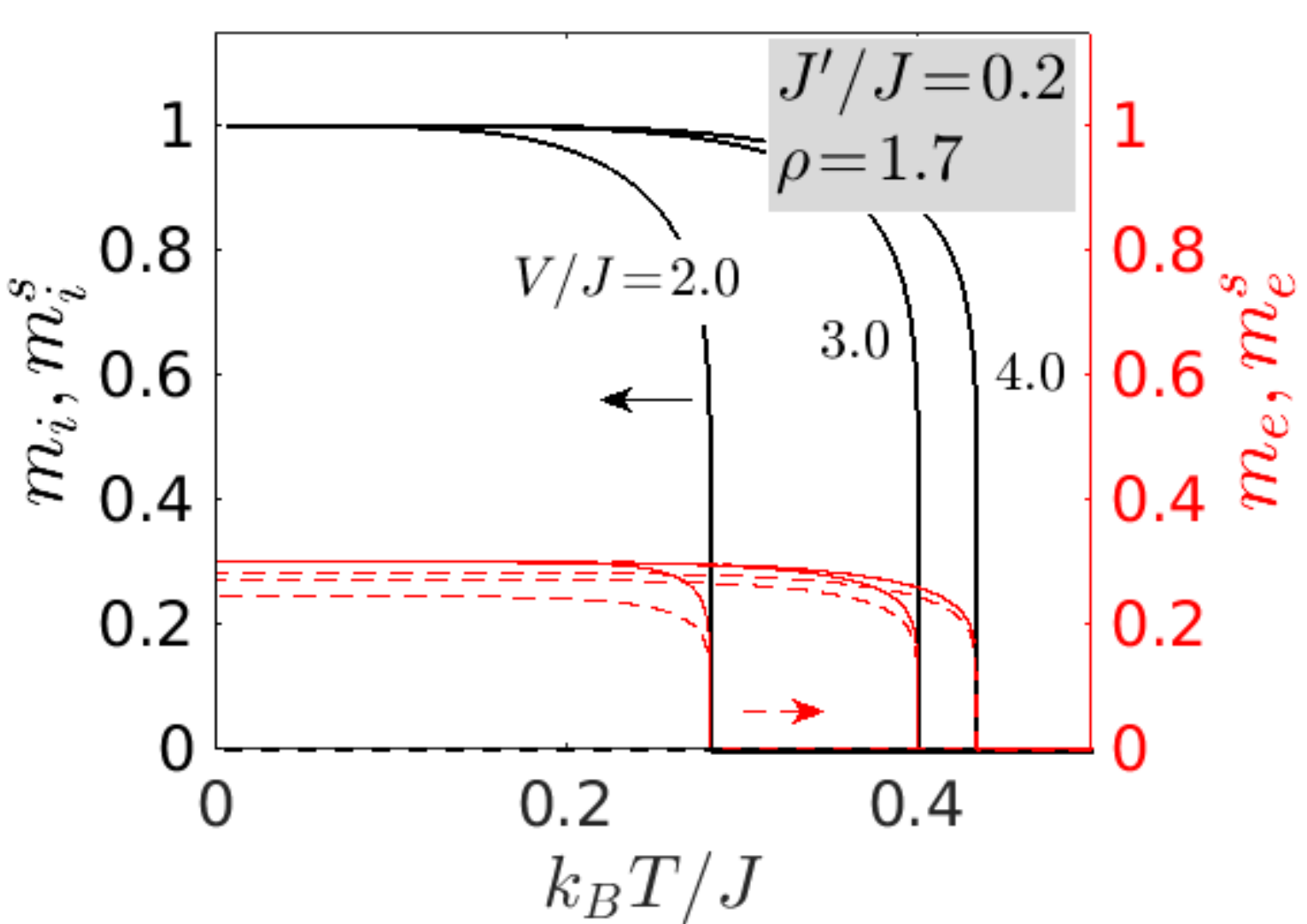}
{\includegraphics[width=0.214\textwidth,height=0.175\textheight,trim=3.5cm 9.3cm 4.7cm 9.8cm, clip]{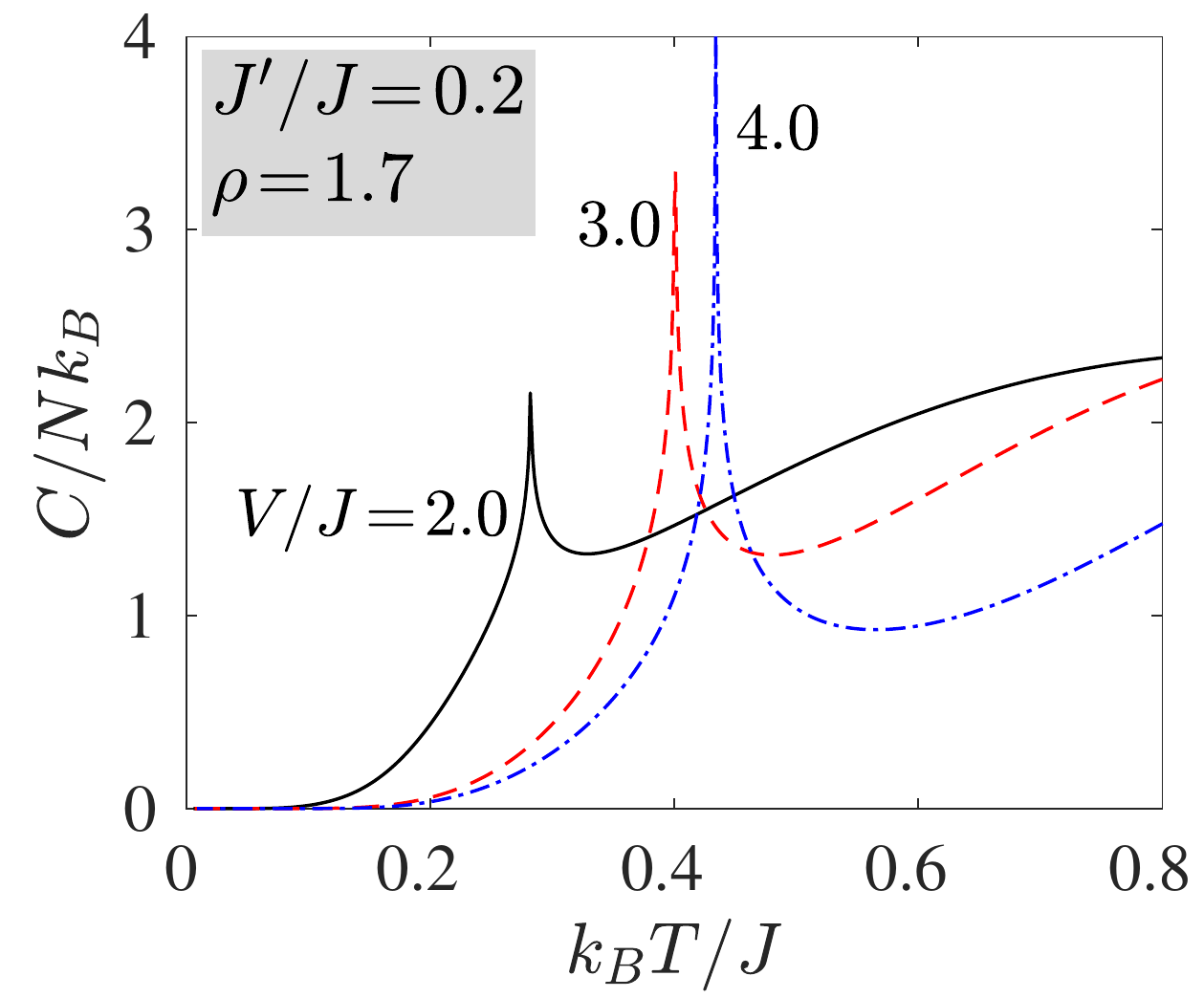}}
\includegraphics[width=0.254\textwidth,height=0.175\textheight,trim=3cm 9.3cm 3cm 9.8cm, clip]{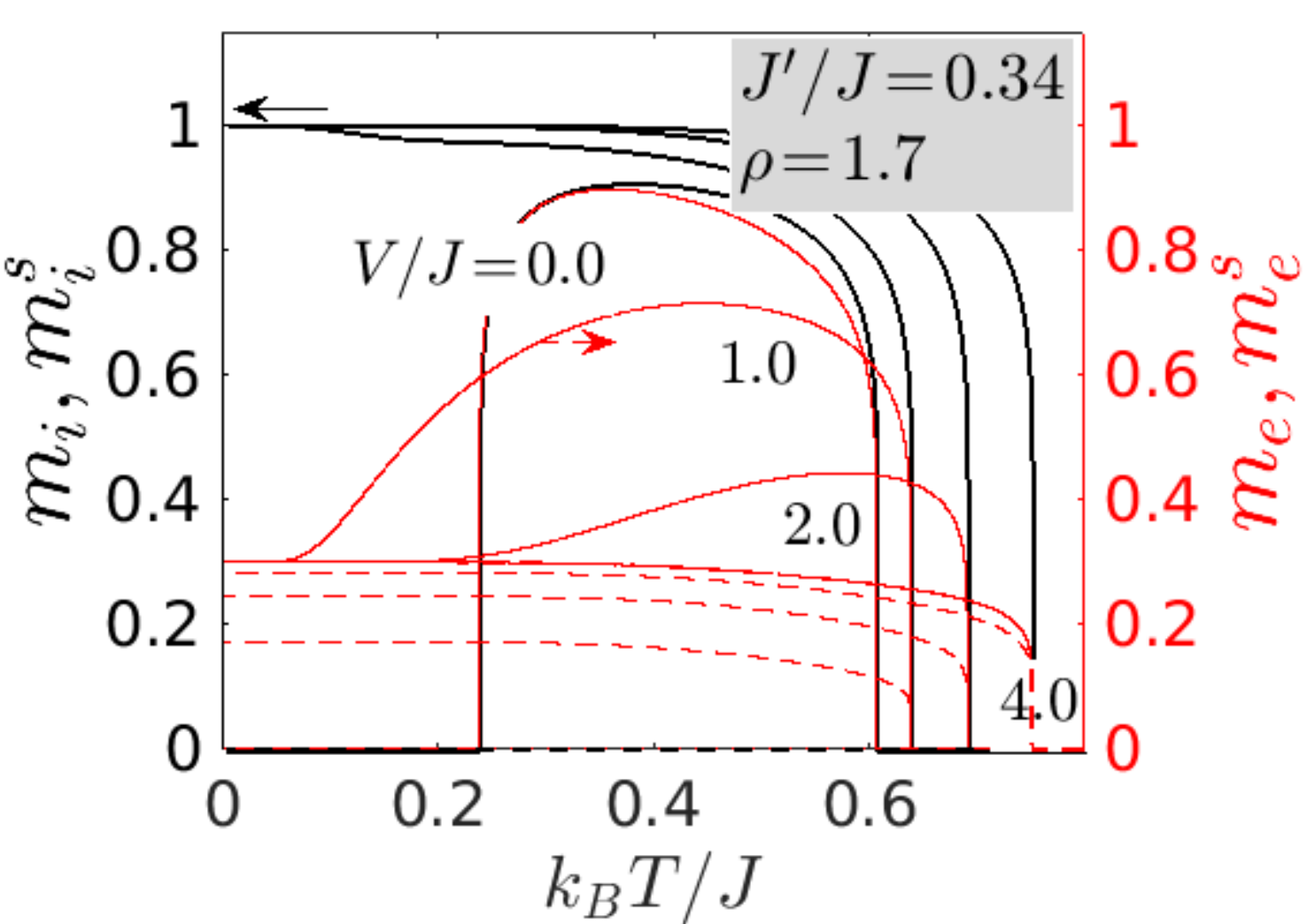}
{\includegraphics[width=0.214\textwidth,height=0.175\textheight,trim=3.5cm 9.3cm 5cm 9.8cm, clip]{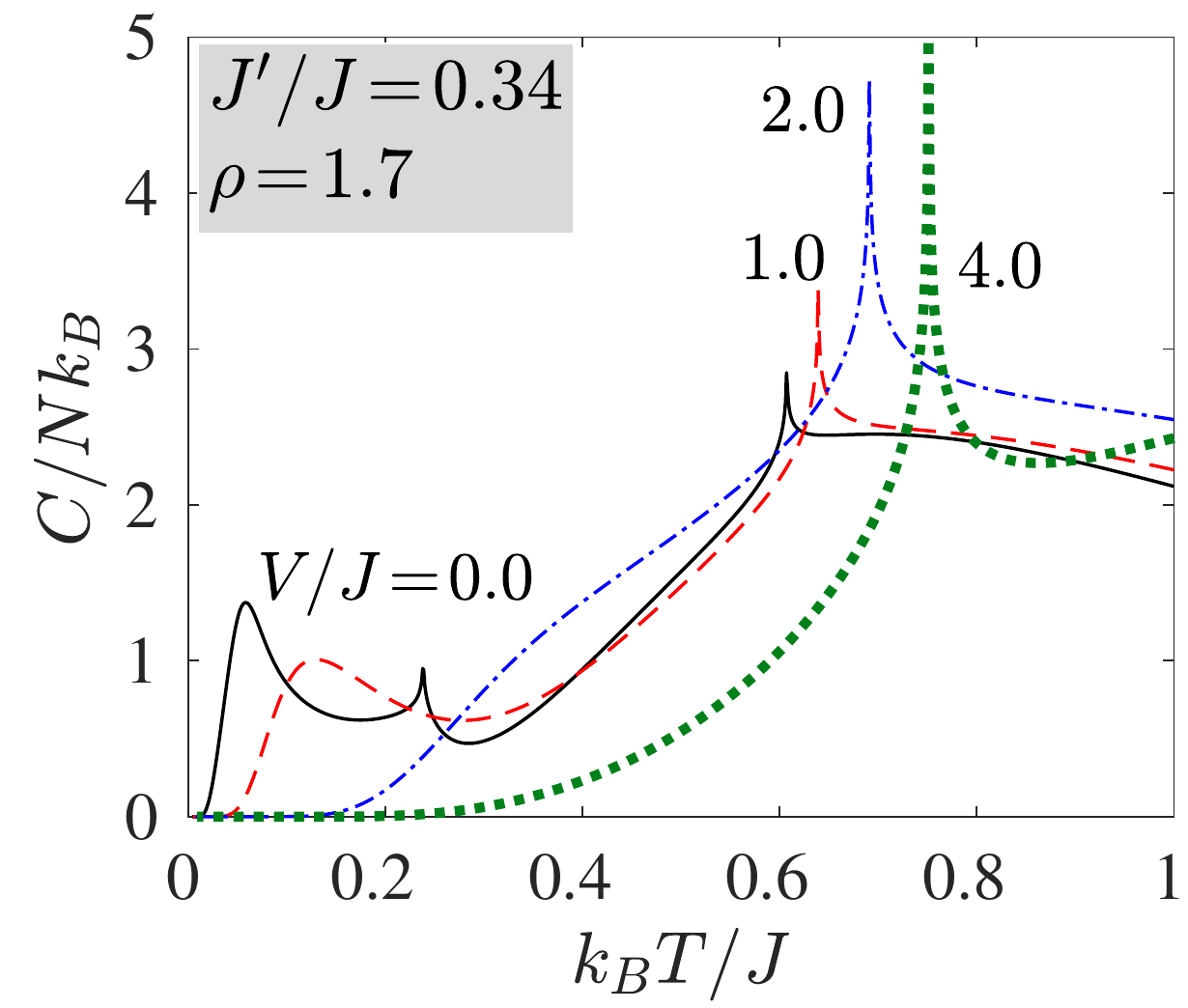}}\\
\includegraphics[width=0.254\textwidth,height=0.175\textheight,trim=3.cm 9.3cm 3.cm 9.8cm, clip]{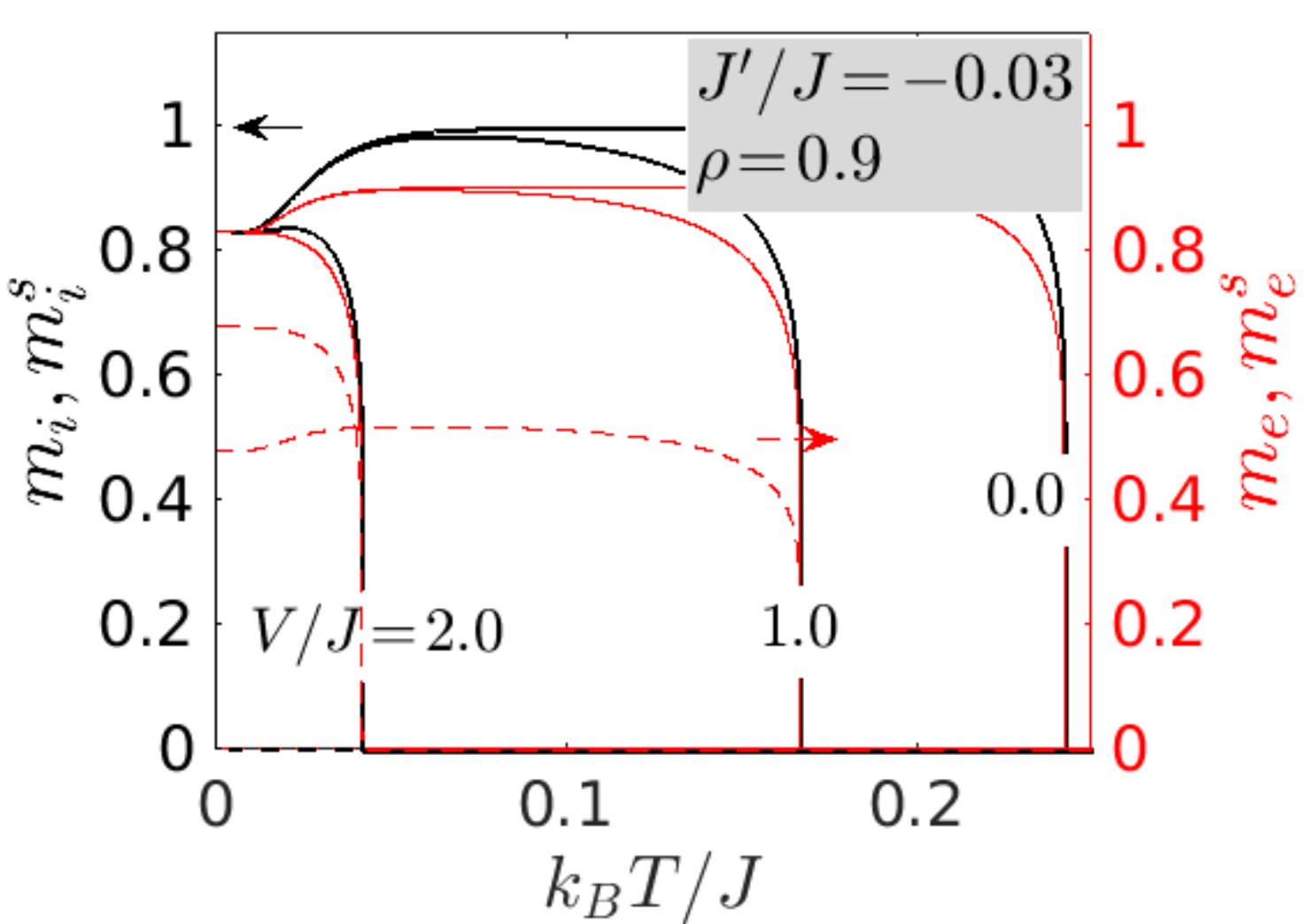}
\includegraphics[width=0.214\textwidth,height=0.175\textheight,trim=3.2cm 9.3cm 4.7cm 9.8cm, clip]{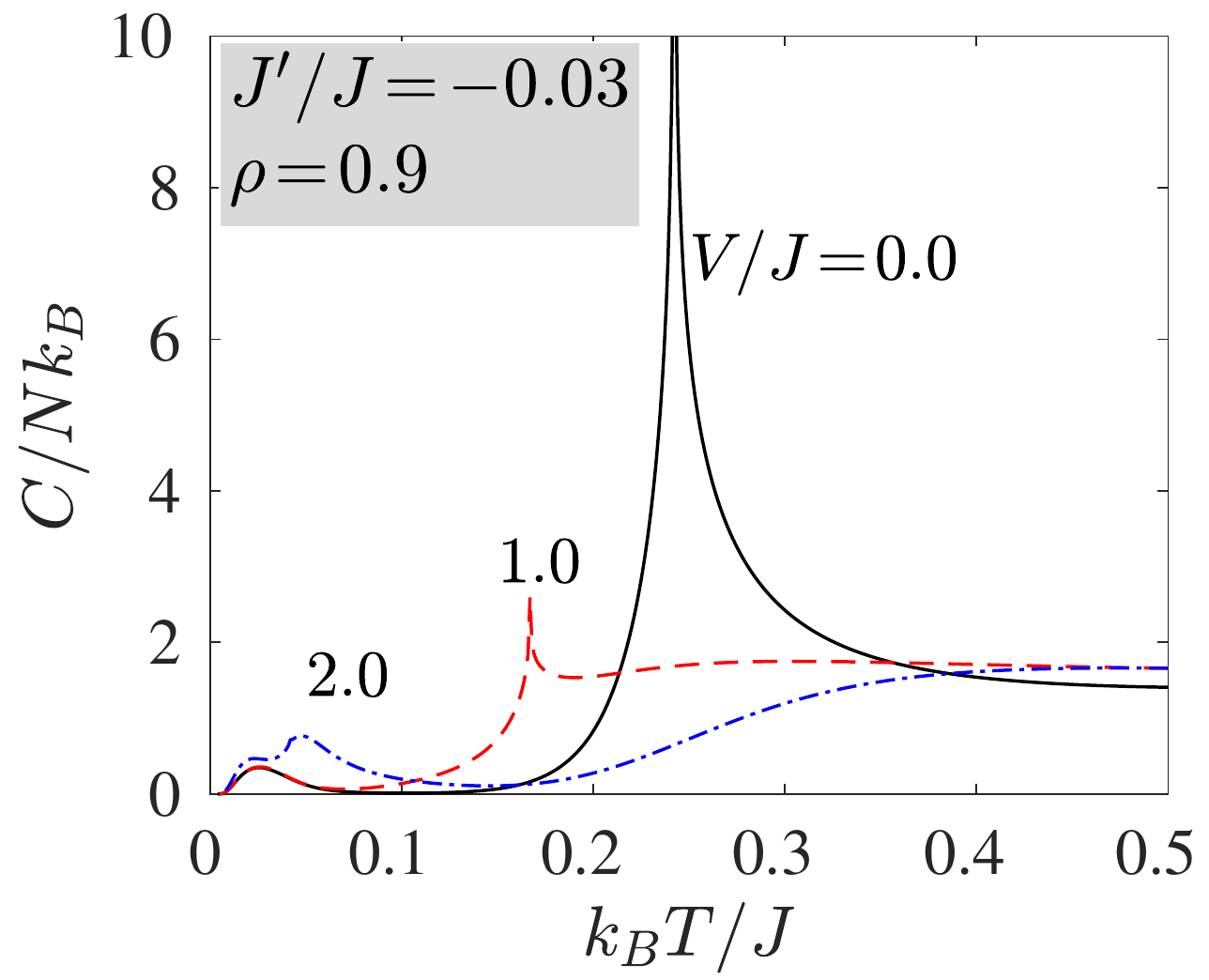}
\includegraphics[width=0.254\textwidth,height=0.175\textheight,trim=3.cm 9.3cm 3.cm 9.8cm, clip]{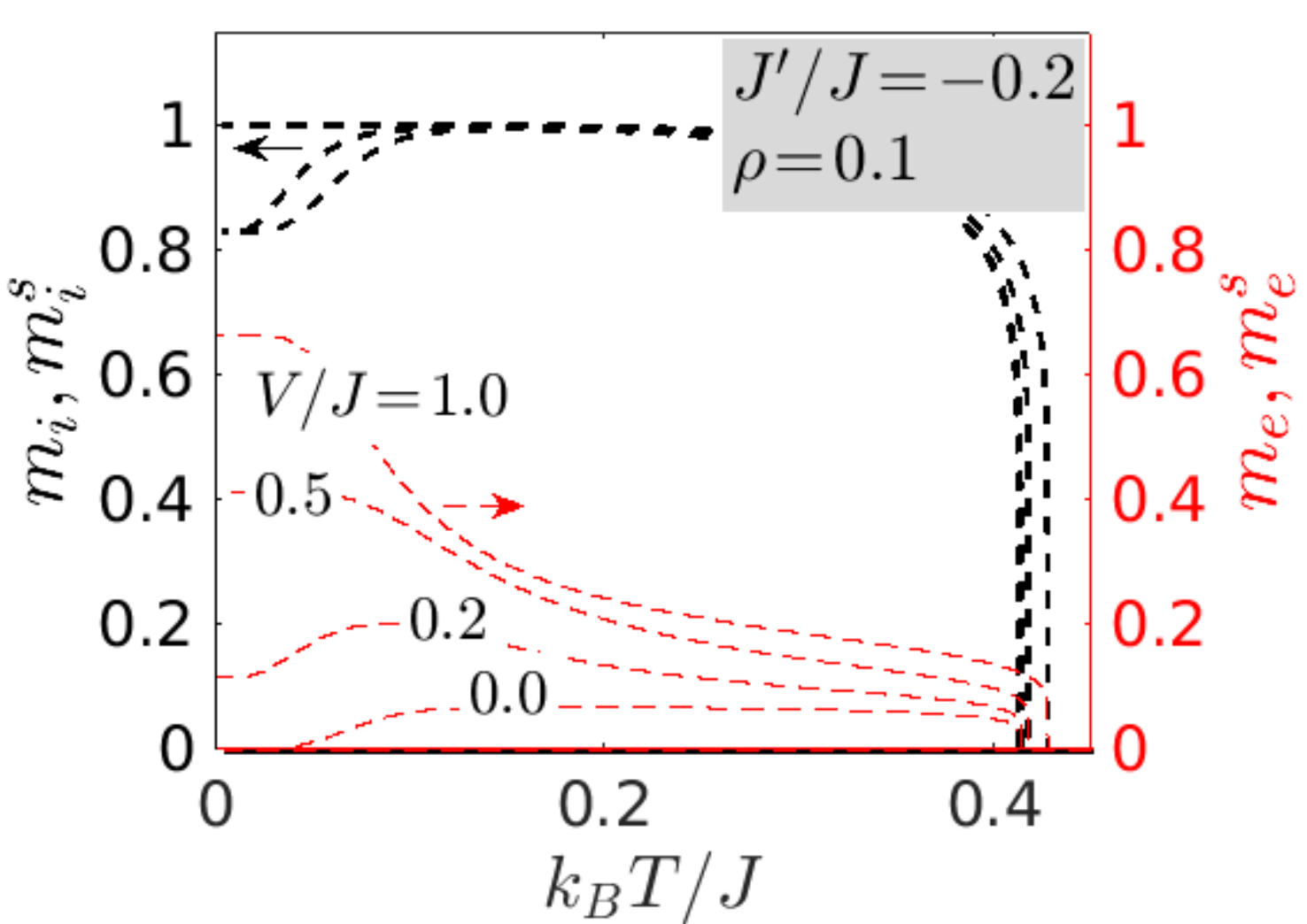}
\includegraphics[width=0.214\textwidth,height=0.175\textheight,trim=3.2cm 9.3cm 4.7cm 9.8cm, clip]{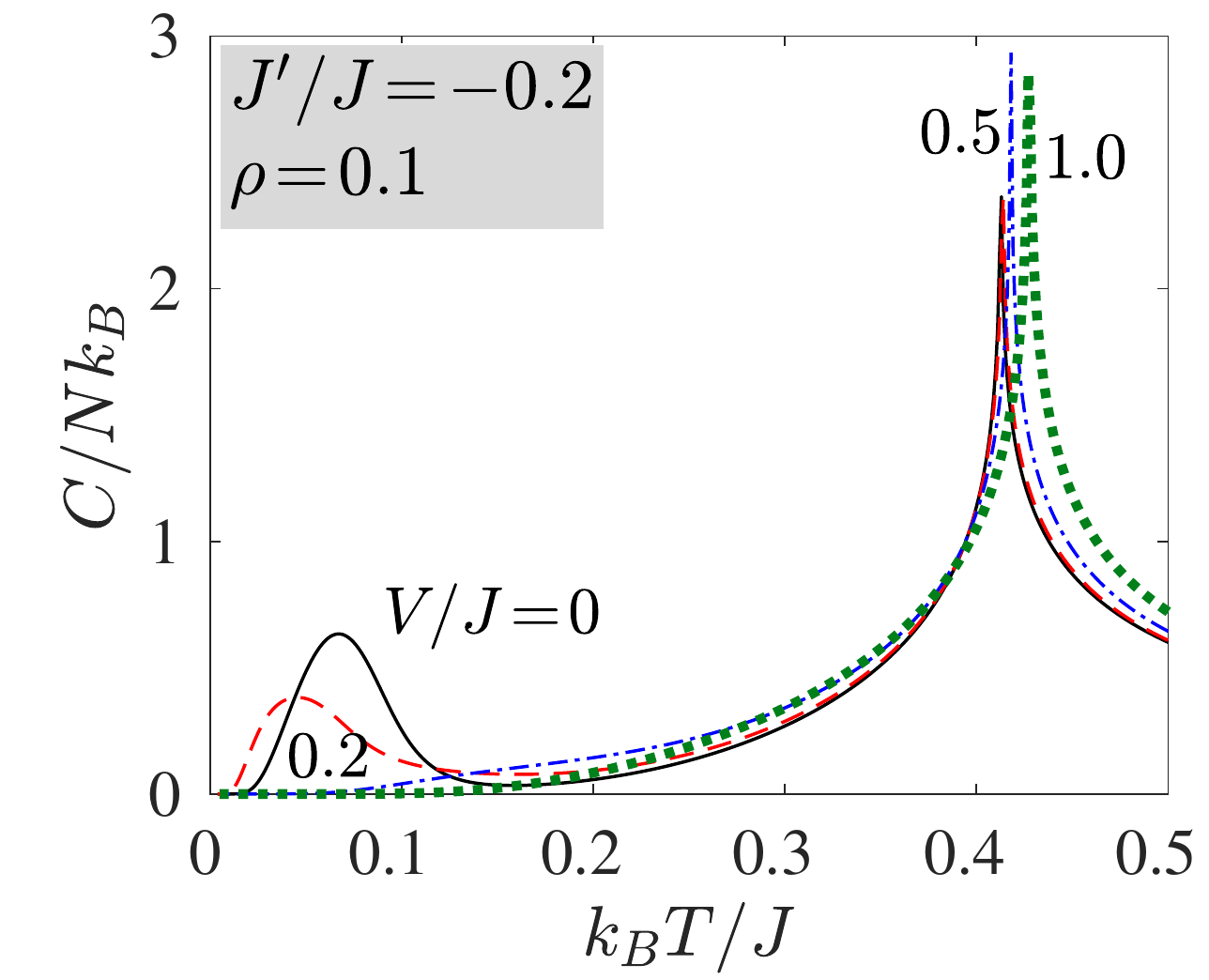}
\caption{\small The  sublattice magnetizations and specific heats as  functions of temperature for the  selected values of $\rho$, $J'/J$ and $V/J$ at $t/J\!=\!1$ and $h/J\!=\!0$ (exact results). The magnetization data shown by solid (dashed) lines illustrate uniform magnetizations $m_i$ and $m_e$ (staggered magnetizations $m_i^s$ and $m_e^s$). The sublattice magnetizations of localized spins ($m_i$ and $m_i^s$) visualized by thick lines are scaled with respect to the left axis, while the sublattice magnetization of the mobile electrons  ($m_e$ and $m_e^s$) are scaled with respect to the right axis.}
\label{fig7}
\end{center}
\end{figure}
 For a very low values of the interaction ratio $|J'|/J$ the spontaneous AF  long-range order is exclusively determined by the spin subsystem, while the staggered magnetization of the electron subsystem vanishes, $m_e^s\!=\!0$. Contrary to this, the AF ordering is present in both spin as well as electron subsystems at the higher values of $|J'|/J$. 
 
\begin{figure}[t!]
\begin{center}
\includegraphics[width=0.254\textwidth,height=0.175\textheight,trim=3.cm 9.3cm 3.cm 9.9cm, clip]{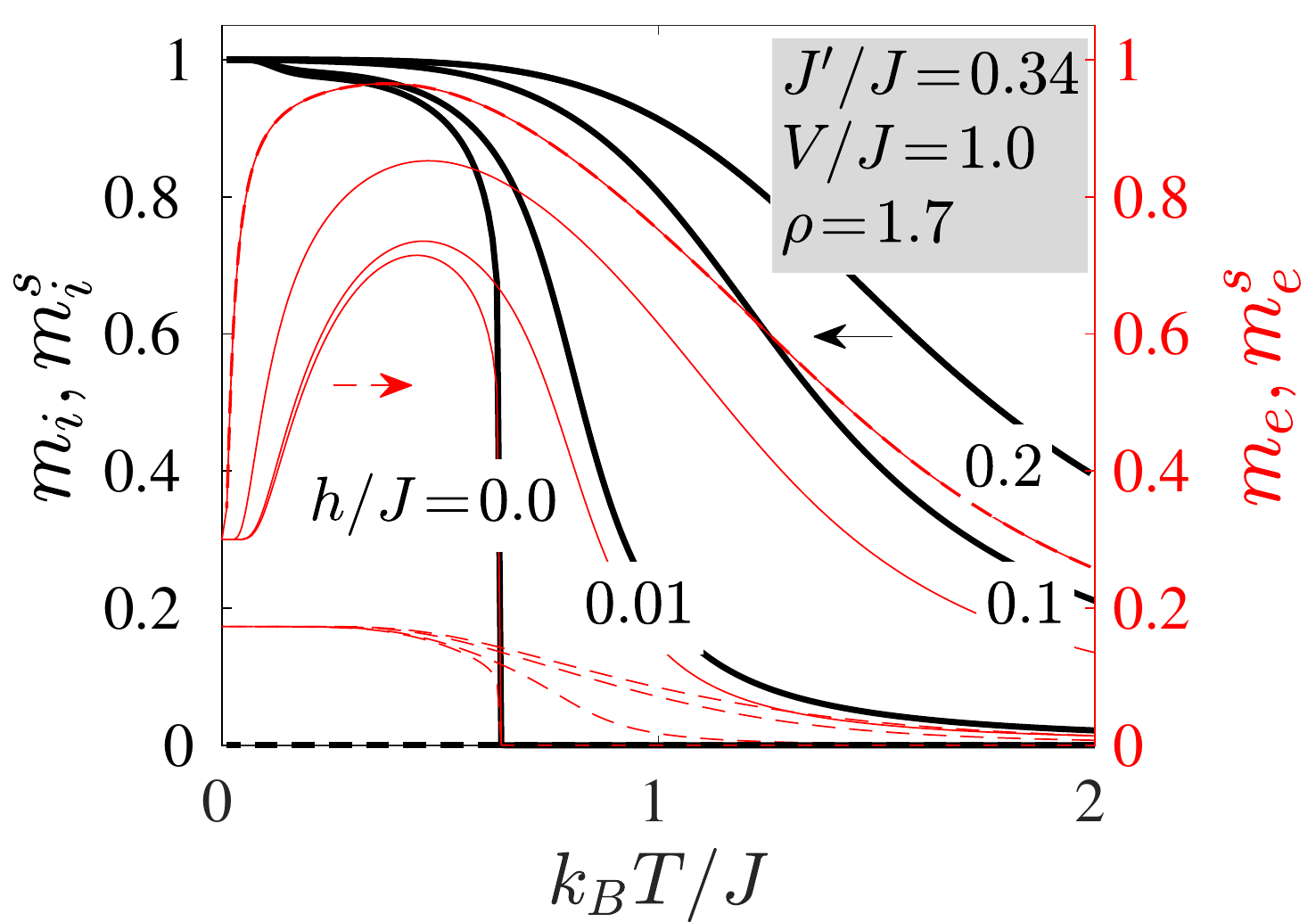}
\includegraphics[width=0.214\textwidth,height=0.175\textheight,trim=3.cm 9.3cm 4.7cm 9.9cm, clip]{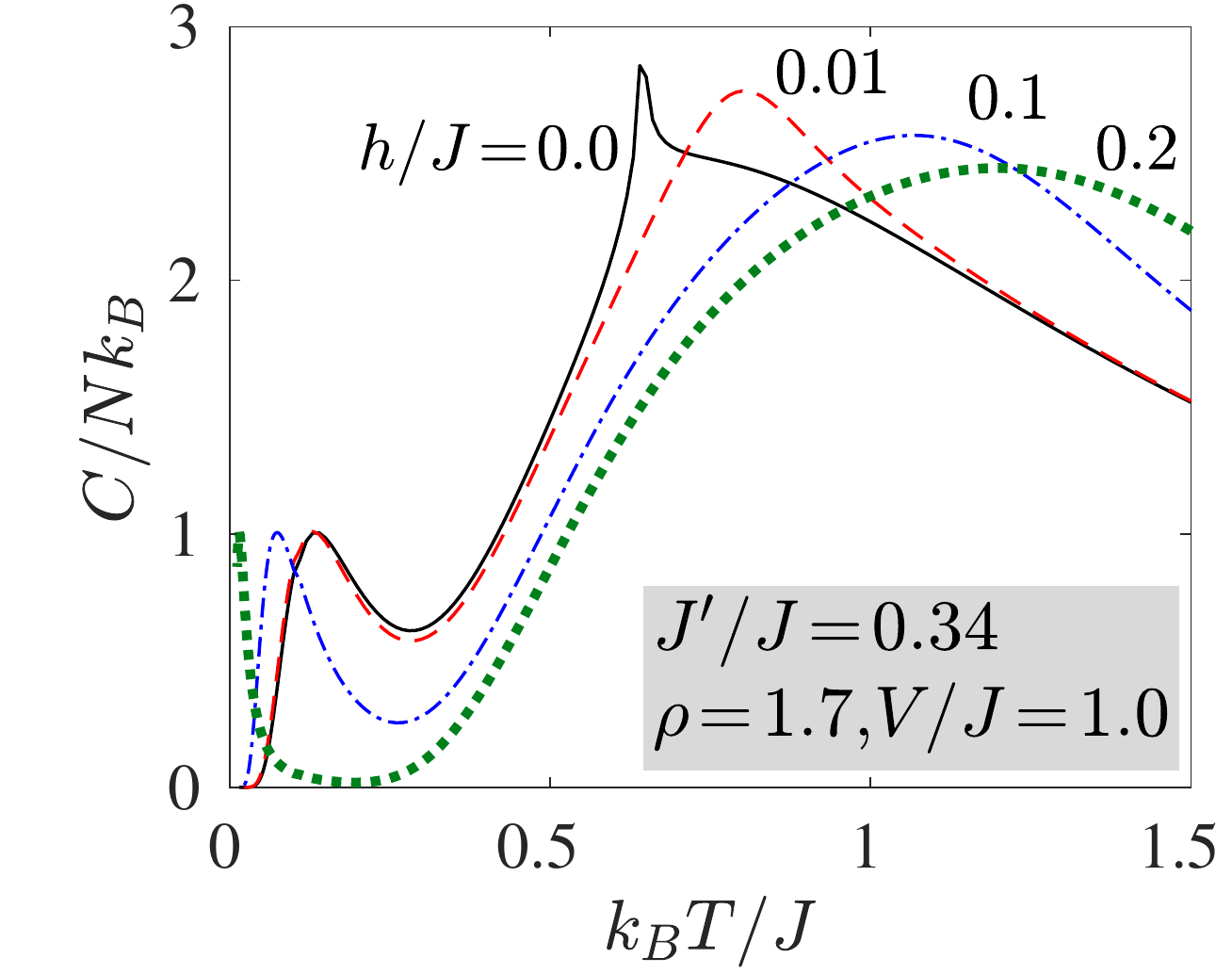}
\includegraphics[width=0.254\textwidth,height=0.175\textheight,trim=3.cm 9.3cm 3cm 9.9cm, clip]{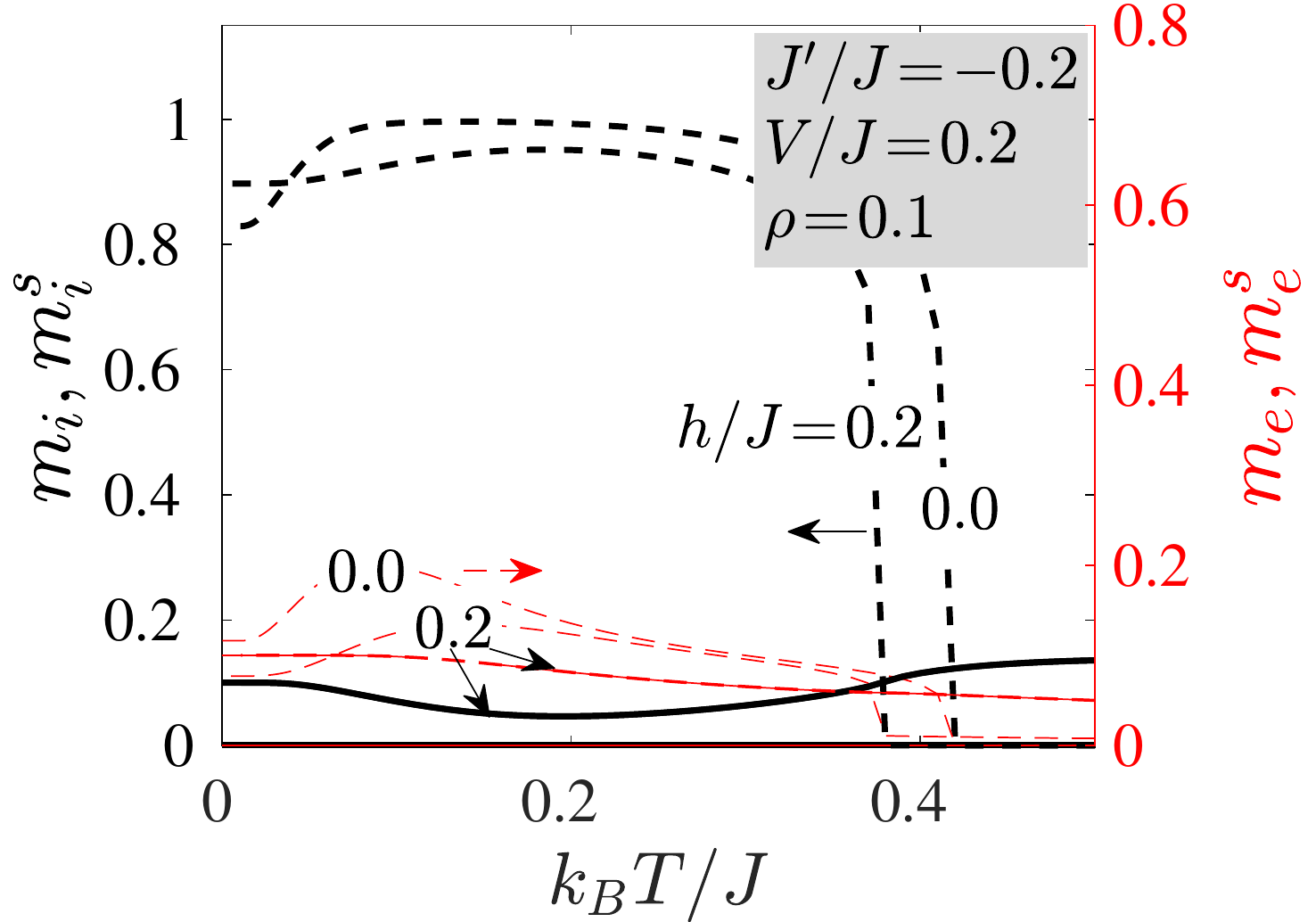}
{\includegraphics[width=0.214\textwidth,height=0.175\textheight,trim=3.cm 9.3cm 4.7cm 9.9cm, clip]{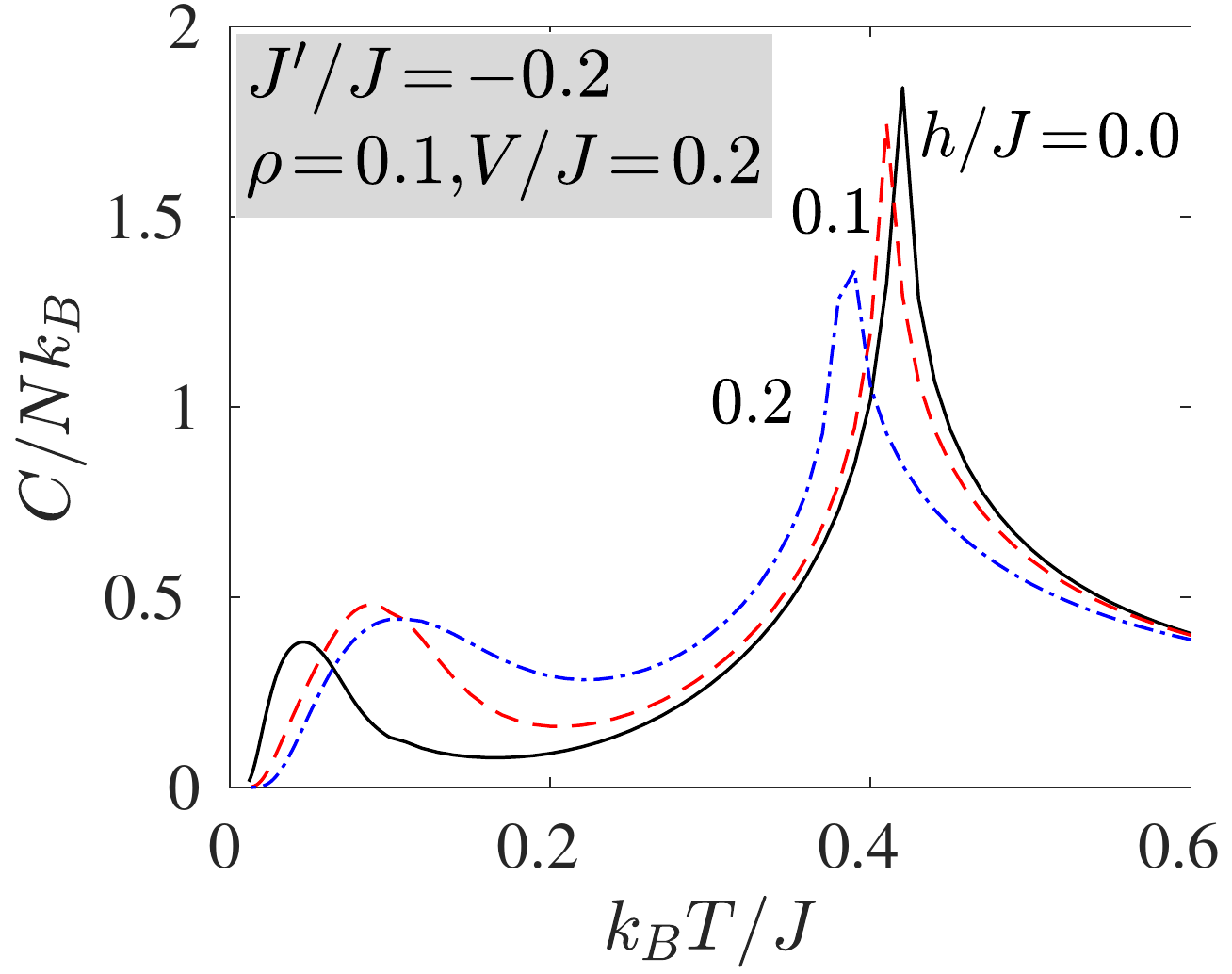}}
\caption{\small   The  sublattice magnetizations and specific heats as  functions of temperature for the selected values of $\rho$, $J'/J$, $V/J$ and $h/J$ at $t/J\!=\!1$. The magnetization data shown by solid (dashed) lines illustrate uniform magnetizations $m_i$ and $m_e$ (staggered magnetizations $m_i^s$ and $m_e^s$). The sublattice magnetizations of localized spins ($m_i$ and $m_i^s$) visualized by thick lines are scaled with respect to the left axis, while the sublattice magnetization of the mobile electrons  ($m_e$ and $m_e^s$) are scaled with respect to the right axis.}
\label{fig8}
\end{center}
\end{figure}

Finally, let us analyze separately the thermal behavior of the sublattice uniform/staggered magnetizations and specific heat (Figs.~\ref{fig7} and~\ref{fig8}) for the non-zero spin-spin interaction $J'/J$. 
Fig.~\ref{fig7} demonstrates how the electric field modifies the magnetic properties of the coupled spin-electron system. 
In principle, the changes in both subsystems could have an equivalent  (cf. for instance $J'/J\!=\!0.2$, $\rho\!=\!1.7$ with $J'/J\!=\!-0.03$, $\rho\!=\!0.9$)  or fully different (cf. for instance $J'/J\!=\!0.34$, $\rho\!=\!1.7$ with $J'/J\!=\!-0.2$, $\rho\!=\!0.1$) character. Moreover, those changes can appear very gradually depending on the model parameters, with a more rapid decline at the critical temperature or even with more significant variations below the critical temperature. As one can expect, the rapid changes in the sublattice magnetizations are reflected in the specific heat as an intriguing low-temperature round maximum.  For the F spin-spin interaction $J'$ the existence of such behaviors can be explained from 
 the huge response of the electron subsystem with respect to thermal fluctuations, which lead to the increment of the uniform magnetization of mobile electrons $m_e$  while leaving the staggered magnetization of the mobile electrons $m_e^s$ unchanged or vice versa for the AF spin-spin interactions.  Similarly, the thermal fluctuations are also responsible for the analogous changes of the uniform and staggered magnetizations of the Ising spins $m_i$ and $m_i^s$, respectively.  It is worth mentioning that the additional magnetic field has a significant impact on the position of this  round maximum as it is illustrated in Fig.~\ref{fig8}. While for the F spin-spin interaction $J'$  the increasing magnetic field reduces its occurrence,  an opposite effect is observed for the AF spin-spin interaction $J'$ when the magnetic field  shifts position of the round maximum to the higher temperatures, and it thus stabilizes the spontaneous magnetic state at the higher values of temperature. 
\section{Conclusion}

In the present work we have studied the coupled spin-electron model on a doubly decorated square lattice using the generalized mapping transformation in combination with the transfer-matrix method or the numerical CTMRG method. Our attention has been primarily focused on clarifying of a simultaneous effect of the applied electric and magnetic fields on a thermal stability of reentrant phase transitions of a coupled spin-electron system on a decorated square lattice. As mentioned previously, the reentrant magnetic transition is a fascinating  phenomenon with a huge application potential. Unfortunately, its occurrence is not trivial as it originates from a competition between various interactions present in a complex multi-component systems. For this reason an exhaustive understanding of mechanisms responsible for its existence is a huge challenge for  researchers. From the technological point of view, reentrant magnetic transitions could be achieved in a certain class of multi-component materials through the chemical doping. Such procedure is however a time consuming and non-economic. 

In the present work we have proposed a new alternative mechanism how to easily and more effectively modulate the magnetic reentrant transitions, using the magnetoelectric effect of the  spin-electron systems. Our analyses clearly demonstrate that the external electric field (with an absence of its magnetic counterpart) may cause a similar effect as the spin-spin interaction.  In addition, it has been found that for sufficiently large electrostatic potential $V/J$ the critical temperature of the order-disorder phase transition may become independent of the electron density with a  fixed value of the critical temperature $k_BT_c/J\!=\!2J'/J\ln(1+\sqrt{2})$ depending solely on the spin-spin interaction. Such property makes the considered spin-electron systems more attractive for applications, because this minimizes their sensitivity on a possible current instability.  In accordance with our assumption it has been confirmed that the applied external fields (the magnetic as well as electric) generally stabilize the F order rather than the AF one. However, the magnetic field can also stabilize the AF order in a narrow parametric space, where it generates novel reentrant magnetic transitions.  Besides, it has been shown that a completely novel reentrance can be realized  exclusively under simultaneous presence of both electric and magnetic fields.  It is worthwhile to remark that the  non-zero electric and magnetic fields have a significant impact on  presence of reentrance upon electron hopping $t$. As a matter of fact, one can either shift or even fully reduce/induce existence of reentrant magnetic transitions varying the value of one or both external fields. 

Finally, we have also analyze in detail the thermal behavior of sublattice uniform/staggered magnetizations. It has been demonstrated that the uniform and staggered magnetization of the spin and electron subsystem exhibit diverse temperature dependencies, whereas significant low-temperature changes resulted in an intriguing round maximum, whose position can be tuned by an applied magnetic field. 

\vspace{0.5cm}
This work was supported by the Slovak Research and Development Agency (APVV) under Grant No. APVV-16-0186 and ERDF EU Grant under contract No. ITMS  26220120047. The financial support provided by the VEGA under Grant No. 1/0043/16 is also gratefully acknowledged. 
%%%%%%%%%%%%%%%%%%%%%%

%%
\end{document}